%% file: neurips_main.tex
\documentclass{article} % For LaTeX2e
\usepackage[preprint]{neurips_2024}

\input{math_commands.tex}

\usepackage{url}
\usepackage{enumitem}
\usepackage{booktabs}
\usepackage{multicol, multirow}
\usepackage{graphicx}
\usepackage{xspace}
\usepackage{times}
\usepackage[hyperindex,breaklinks,colorlinks,citecolor=darkblue]{hyperref} % allow link break (for arxiv) 
\usepackage{wrapfig}
\usepackage{textcomp} % For special characters
\usepackage{hyperref}
\usepackage{caption}
\usepackage{subcaption}
\usepackage{todonotes}
\usepackage{listings}
\usepackage{xcolor}
\usepackage{tcolorbox}

\definecolor{myblue}{rgb}{0.0, 0.0, 0.5}
\definecolor{mygray}{rgb}{0.95, 0.95, 0.95}

\newtcolorbox{codebox}[1][]{colback=blue!5!white, colframe=myblue, fonttitle=\bfseries, title=#1}

\definecolor{darkblue}{rgb}{0.0, 0.0, 0.55}
\definecolor{myblue}{rgb}{0,0.45,0.74}
\definecolor{myred}{rgb}{0.85,0.33,0.1}
\definecolor{deepgreen}{RGB}{0, 150, 0}

\title{\framework: Bridging Large Language Models and Code Repositories via Code Graph Databases}

\author{Xiangyan Liu$^{1,3,*}$ \quad Bo Lan$^{2,*}$ \quad Zhiyuan Hu$^{1}$ \quad \textbf{Yang Liu$^{3}$} \quad \textbf{Zhicheng Zhang$^{3}$} \\
\textbf{Fei Wang$^{2}$} \quad \textbf{Michael Shieh$^{1,\dagger}$} \quad \textbf{Wenmeng Zhou$^{3}$} \\
$^1$ National University of Singapore \quad
$^2$ Xi'an Jiaotong University \quad
$^3$ Alibaba Group \\
\texttt{\{liu.xiangyan@u.nus.edu, bolan@stu.xjtu.edu.cn\}}
}

\newcommand{\framework}{\textsc{CodexGraph}\xspace}

\begin{document}

\maketitle
\renewcommand{\thefootnote}{\fnsymbol{footnote}}
\footnotetext[1]{Equal contribution. Work done during Xiangyan's internship at Alibaba.}
\footnotetext[2]{Corresponding author: \texttt{michaelshieh@comp.nus.edu.sg}}
\renewcommand{\thefootnote}{\arabic{footnote}}

\begin{abstract}
Large Language Models (LLMs) excel in stand-alone code tasks like HumanEval and MBPP, but struggle with handling entire code repositories.
This challenge has prompted research on enhancing LLM-codebase interaction at a repository scale.
Current solutions rely on similarity-based retrieval or manual tools and APIs, each with notable drawbacks.
Similarity-based retrieval often has low recall in complex tasks, while manual tools and APIs are typically task-specific and require expert knowledge, reducing their generalizability across diverse code tasks and real-world applications.
To mitigate these limitations, we introduce \framework, a system that integrates LLM agents with graph database interfaces extracted from code repositories.
By leveraging the structural properties of graph databases and the flexibility of the graph query language, \framework enables the LLM agent to construct and execute queries, allowing for precise, code structure-aware context retrieval and code navigation.
We assess \framework using three benchmarks: CrossCodeEval, SWE-bench, and EvoCodeBench. Additionally, we develop five real-world coding applications.
With a unified graph database schema, \framework demonstrates competitive performance and potential in both academic and real-world environments, showcasing its versatility and efficacy in software engineering.
Our application demo: \url{https://github.com/modelscope/modelscope-agent/tree/master/apps/codexgraph_agent}.

\end{abstract}

\section{Introduction}
\vspace{-2mm}
\begin{figure}[!t]
    \centering
    \vspace{-2em}
    \includegraphics[width=0.9\textwidth]{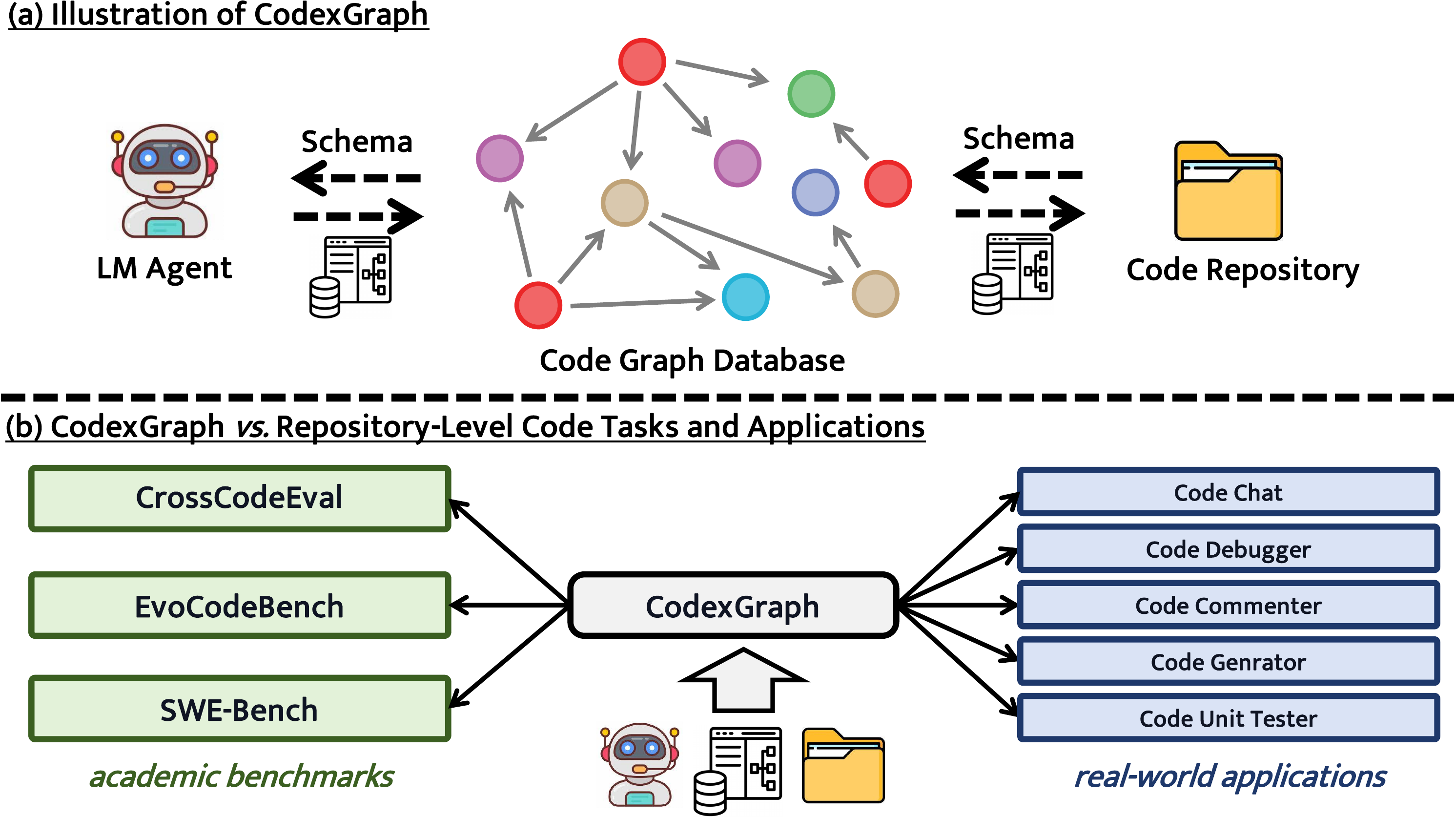}
    \vspace{-2mm}
    \caption{\small (a) Using a unified schema, \framework employs code graph databases as interfaces that allow LLM agents to interact seamlessly with code repositories. (b) \framework supports the management of a wide range of tasks, from academic-level code benchmarks to real-world software engineering applications.}
    \label{fig:motivation}
    \vspace{-2.3em}
\end{figure}
Large Language Models (LLMs) excel in code tasks, impacting automated software engineering \citep{chen2021evaluating, aider, yang2024swe, opendevin2024}. Repository-level tasks \citep{zhang2023repocoder, jimenez2023swe, ding2024crosscodeeval, li2024evocodebench} mimic software engineers' work with large codebases \citep{kovrigin2024importance}. These tasks require models to handle intricate dependencies and comprehend project structure \citep{jiang2024survey, sun2024survey}.

Current LLMs struggle with long-context inputs, limiting their effectiveness with large codebases \citep{jimenez2023swe} and lengthy sequences reasoning \citep{liu2024lost}. Researchers have proposed methods to enhance LLMs by retrieving task-relevant code snippets and structures, improving performance in complex software development \citep{deng2024r2c2, arora2024masai, ma2024understand}.
However, these approaches mainly rely on either similarity-based retrieval \citep{jimenez2023swe, cheng2024dataflow, liu2024graphcoder} or manual tools and APIs \citep{zhang2024autocoderover, moatlesstools}. Similarity-based retrieval methods, common in Retrieval-Augmented Generation (RAG) systems \citep{lewis2020retrieval}, often struggle with complex reasoning for query formulation \citep{jimenez2023swe} and handling intricate code structures \citep{phan2024repohyper}, leading to low recall rates.
Meanwhile, existing tool/API-based interfaces that connect codebases and LLMs are typically task-specific and require extensive expert knowledge \citep{moatlesstools, chen2024coder}.
Furthermore, our experimental results in Section \ref{exp:results} indicate that the two selected methods lack flexibility and generalizability for diverse repository-level code tasks. \looseness=-1

Recent studies have demonstrated the effectiveness of graph structures in code repositories \citep{phan2024repohyper, cheng2024dataflow}.
Meanwhile, inspired by recent advances in graph-based RAG \citep{edge2024local, liu2024graphcoder, he2024g} and the application of executable code (such as SQL, Cypher, and Python) to consolidate LLM agent actions \citep{wang2024executable, li2024unioqa, xue2023db}, we present \framework, as shown in Figure \ref{fig:motivation} (a).
\framework alleviates the limitations of existing approaches by bridging code repositories with LLMs through graph databases. \framework utilizes static analysis to extract code graphs from repositories using a task-agnostic schema that defines the nodes and edges within the code graphs.
In these graphs, nodes represent source code symbols such as \texttt{MODULE}, \texttt{CLASS}, and \texttt{FUNCTION}, and each node is enriched with relevant meta-information.
The edges between nodes represent the relationships among these symbols, such as \texttt{CONTAINS}, \texttt{INHERITS}, and \texttt{USES} (see Figure \ref{fig:schema} for an illustrative example).
By leveraging the structural properties of graph databases, \framework enhances the LLM agent's comprehension of code structures.
\framework leverages repository code information and graph structures for global analysis and multi-hop reasoning, enhancing code task performance. When users provide code-related inputs, the LLM agent analyzes the required information from the code graphs, constructs flexible queries using graph query language, and locates relevant nodes or edges. This enables precise and efficient retrieval, allowing for effective scaling to larger repository tasks.

To evaluate the effectiveness of the \framework, we assess its performance across \textbf{three} challenging and representative repository-level benchmarks: CrossCodeEval \citep{ding2024crosscodeeval}, SWE-bench \citep{yang2024swe} and EvoCodeBench \citep{li2024evocodebench}.
Our experimental results demonstrate that, by leveraging a unified graph database schema (Section \ref{build_code_graph_db}) and a simple workflow design (Section \ref{LLM_interacts_with_code_graph_db}), the \framework achieves competitive performance across all academic benchmarks, especially when equipped with more advanced LLMs.
Furthermore, as illustrated in Figure \ref{fig:motivation} (b), to address real-world software development needs, we extend \framework to the feature-rich ModelScope-Agent \citep{li2023modelscopeagent} framework.
Section \ref{sec:application} highlights \textbf{five} real-world application scenarios, including code debugging and writing code comments, showcasing the versatility and efficacy of \framework in practical software engineering tasks.
% Refer to Figure \ref{fig:motivation} for a visual illustration.

\textbf{Our contributions are from three perspectives:}
\begin{itemize}[leftmargin=12pt, nosep]
\item \textbf{Pioneering code retrieval system:} We introduce \framework, integrating code repositories with LLMs via graph databases for enhanced code navigation and understanding.
\item \textbf{Benchmark performance:} We demonstrate \framework's competitive performance on three challenging and representative repository-level code benchmarks.
\item \textbf{Practical applications:} We showcase \framework's versatility in five real-world software engineering scenarios, proving its value beyond academic settings.
\end{itemize}

\section{Related Work}
\subsection{Repository-Level Code Tasks}
Repository-level code tasks have garnered significant attention due to their alignment with real-world production environments \citep{bairi2023codeplan, luo2024repoagent, devin2024, kovrigin2024importance}.
Unlike traditional standalone code-related tasks such as HumanEval \citep{chen2021evaluating} and MBPP \citep{austin2021program}, which often fail to capture the complexities of real-world software engineering, repository-level tasks necessitate models to understand cross-file code structures and perform intricate reasoning \citep{liu2024graphcoder,ma2024understand, sun2024survey}.
These sophisticated tasks can be broadly classified into two lines of work based on their inputs and outputs.
The first line of work involves natural language to code repository tasks, exemplified by benchmarks like DevBench \citep{li2024devbench} and SketchEval \citep{zan2024codes}, where models generate an entire code repository from scratch based on a natural language description of input requirements.
State-of-the-art solutions in this area often employ multi-agent frameworks such as ChatDev \citep{qian2023communicative} and MetaGPT \citep{hong2023metagpt} to handle the complex process of generating a complete codebase.
The second line of work, which our research focuses on, includes tasks that integrate both a natural language description and a reference code repository, requiring models to perform tasks like repository-level code completion \citep{zhang2023repocoder, shrivastava2023repofusion, liu2023repobench, ding2024crosscodeeval, su2024arks}, automatic GitHub issue resolution \citep{jimenez2023swe}, and repository-level code generation \citep{li2024evocodebench}.
To assess the versatility and effectiveness of our proposed system \framework, we evaluate it on three diverse and representative benchmarks including CrossCodeEval \citep{ding2024crosscodeeval} for code completion, SWE-bench \citep{jimenez2023swe} for Github issue resolution, and EvoCodeBench \citep{li2024evocodebench} for code generation. \looseness=-1

\subsection{Retrieval-Augmented Code Generation}
\label{related_work:RACG}
Retrieval-Augmented Generation (RAG) systems primarily aim to retrieve relevant content from external knowledge bases to address a given question, thereby maintaining context efficiency while reducing hallucinations in private domains \citep{lewis2020retrieval, shuster2021retrieval}.
For repository-level code tasks, which involve retrieving and manipulating code from repositories with complex dependencies, RAG systems—referred to here as Retrieval-Augmented Code Generation (RACG) \citep{jiang2024survey}—are utilized to fetch the necessary code snippets or code structures from the specialized knowledge base of code repositories.
Current RACG methodologies can be divided into three main paradigms: the first paradigm involves similarity-based retrieval, which encompasses term-based sparse retrievers \citep{, stephen09bm25, jimenez2023swe} and embedding-based dense retrievers \citep{guo2022unixcoder, zhang2023repocoder}, with advanced approaches integrating structured information into the retrieval process \citep{phan2024repohyper, cheng2024dataflow, liu2024graphcoder}.
The second paradigm consists of manually designed code-specific tools or APIs that rely on expert knowledge to create interfaces for LLMs to interact with code repositories for specific tasks \citep{zhang2024autocoderover, deshpande2024class, arora2024masai}.
The third paradigm combines both similarity-based retrieval and code-specific tools or APIs \citep{moatlesstools}, leveraging the reasoning capabilities of LLMs to enhance context retrieval from code repositories.
Apart from the three paradigms, Agentless \citep{xia2024agentless} preprocesses the code repository's structure and file skeleton, allowing the LLMs to interact with the source code.
Our proposed framework, \framework, aligns most closely with the second paradigm but distinguishes itself by discarding the need for expert knowledge and task-specific designs.
By using code graph databases as flexible and universal interfaces, which also structurally store information to facilitate the code structure understanding of LLMs, \framework can navigate the code repositories and manage multiple repository-level code tasks, providing a versatile and powerful solution for RACG.

\section{\framework: Enable LLMs to Navigate the Code Repository}
\framework is a system that bridges code repositories and large language models (LLMs) through code graph database interfaces.
It indexes input code repositories using static analysis, storing code symbols and relationships as nodes and edges in a graph database according to a predefined schema.
When presented with a coding question, \framework leverages the LLM agent to generate graph queries, which are executed to retrieve relevant code fragments or code structures from the database.
The detailed processes of constructing the code graph database and the LLM agent's interactions with it are explained in sections \ref{build_code_graph_db} and \ref{LLM_interacts_with_code_graph_db}, respectively.

\subsection{Build Code Graph Database from Repository Codebase} \label{build_code_graph_db}

\begin{figure}[!t]
    \centering
    \includegraphics[width=0.95\textwidth]{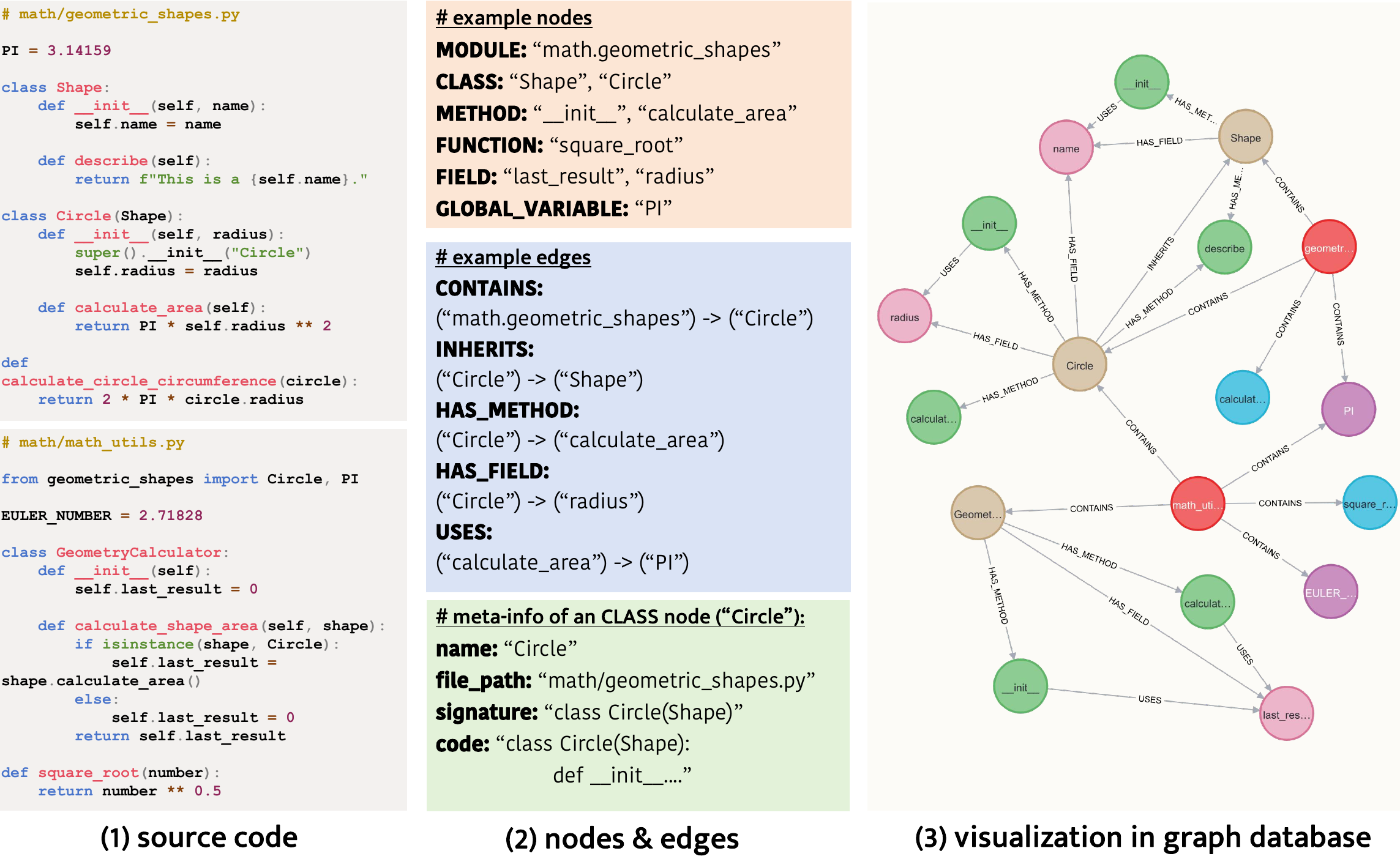}
    \caption{\small Illustration of the process for indexing source code to generate a code graph based on the given graph database schema. Subfigure (3) provides a visualization example of the resultant code graph in Neo4j.}
    \label{fig:schema}
\end{figure}

\paragraph{Schema.}
We abstract code repositories into code graphs where nodes represent symbols in the source code, and edges represent relationships between these symbols.
The schema defines the types of nodes and edges, directly determining how code graphs are stored in the graph database.
Different programming languages typically require different schemas based on their characteristics.
In our project, we focus on Python and have empirically designed a schema tailored to its features, with node types including \texttt{MODULE}, \texttt{CLASS}, \texttt{METHOD}, \texttt{FUNCTION}, \texttt{FIELD}, and \texttt{GLOBAL\_VARIABLE}, and edge types including \texttt{CONTAINS}, \texttt{INHERITS}, \texttt{HAS\_METHOD}, \texttt{HAS\_FIELD}, and \texttt{USES}.\looseness=-1

Each node type has corresponding attributes to represent its meta-information.
For instance, \texttt{METHOD} nodes have attributes such as \texttt{name}, \texttt{file\_path}, \texttt{class}, \texttt{code}, and \texttt{signature}.
For storage efficiency, nodes with a code attribute do not store the code snippet directly in the graph database but rather an index pointing to the corresponding code fragment.
Figure \ref{fig:schema} illustrates a sample code graph derived from our schema, and Appendix \ref{app:schema} shows the details of the schema.

\begin{figure}[!t]
    \centering
    \includegraphics[width=0.95\textwidth]{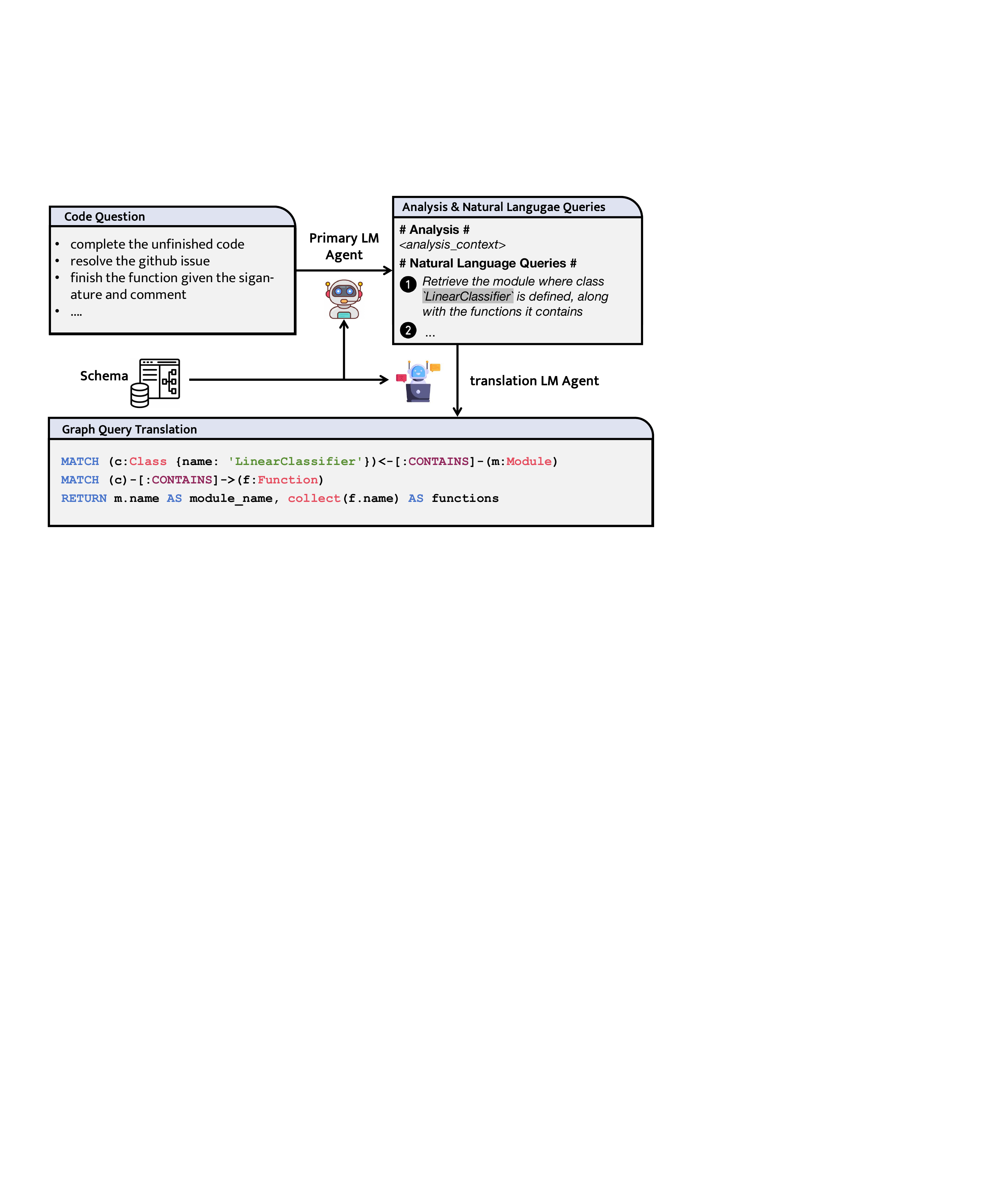}
    \caption{\small The primary LLM agent analyzes the given code question, \textbf{writting} natural language queries. These queries are then processed by the translation LLM agent, which \textbf{translates} them into executable graph queries.}
    \label{fig:translation}
\end{figure}

\paragraph{Phase 1: Shallow indexing.} 
The code graph database construction process consists of two phases, beginning with the input of the code repository and schema.
The first phase employs a shallow indexing method, inspired by Sourcetrail's static analysis process \footnote{\url{https://github.com/CoatiSoftware/Sourcetrail}}, to perform a single-pass scan of the entire repository.
During this scan, symbols and relationships are extracted from each Python file, processed only once, and stored as nodes and edges in the graph database.
Concurrently, meta-information for these elements is recorded.
This approach ensures speed and efficiency, capturing all nodes and their meta-information in one pass.
However, the shallow indexing phase has limitations due to its single-pass nature.
Some important edges, particularly certain \texttt{INHERITS} and \texttt{CONTAINS} relationships, may be overlooked as they might require context from multiple files.

\paragraph{Phase 2: Complete the edges.}
The second phase addresses the limitations of shallow indexing by focusing on cross-file relationships.
We employ Depth-First Search (DFS) to traverse each code file, using abstract syntax tree parsing to identify modules and classes.
This approach is particularly effective in resolving Python's re-export issues.
We convert relative imports to absolute imports, enabling accurate establishment of cross-file \texttt{CONTAINS} relationships through graph queries.
Simultaneously, we record \texttt{INHERITS} relationships for each class.
For complex cases like multiple inheritance, DFS is used to establish edges for inherited \texttt{FIELD} and \texttt{METHOD} nodes within the graph database.
This comprehensive approach ensures accurate capture of both intra-file and cross-file relationships, providing a complete representation of the codebase structure.

\paragraph{Summary.}
Our code graph database design offers four key advantages for subsequent use.
\emph{First}, it ensures efficient storage by storing code snippets as indexed references rather than directly in the graph database.
\emph{Second}, it enables multi-granularity searches, from module-level to variable-level, accommodating diverse analytical needs.
\emph{Third}, it facilitates topological analysis of the codebase, revealing crucial insights into hierarchical and dependency structures.
\emph{Last}, this schema design supports multiple tasks without requiring modifications, demonstrating its versatility and general applicability.
These features collectively enhance the system's capability to handle complex code analysis tasks effectively across various scenarios.

\subsection{Large Language Models Interaction with Code Graph Database}
\label{LLM_interacts_with_code_graph_db}
\paragraph{Code structure-aware search.}
\framework leverages the flexibility of graph query language to construct complex and composite search conditions.
By combining this flexibility with the structural properties of graph databases, the LLM agent can effectively navigate through various nodes and edges in the code graph.
This capability allows for intricate queries such as: \textit{``Find classes under a certain module that contain a specific method''}, or \textit{``Retrieve the module where a certain class is defined, along with the functions it contains''}.
This approach enables code structure-aware searches, providing a level of code retrieval that is difficult to achieve with similarity-based retrieval methods \citep{stephen09bm25, guo2022unixcoder} or conventional code-specific tools and APIs \citep{zhang2024autocoderover, deshpande2024class}.

\paragraph{Write then translate.}
LLM agents are powered by LLMs and operate based on user-provided prompts to break down tasks, utilize tools, and perform reasoning.
This design is effective for handling specific, focused tasks \citep{gupta2022visprog, yuan2024mora}, but when tasks are complex and multifaceted, LLM agents may underperform.
This limitation has led to the development of multi-agent systems \citep{hong2023metagpt, qian2023communicative, guo2024survey}, where multiple LLM agents independently handle parts of the task.
Inspired by this approach, \framework implements a streamlined ``write then translate'' strategy to optimize LLM-database interactions.

As illustrated in Figure \ref{fig:translation}, the primary LLM agent focuses on understanding context and generating natural language queries based on the user's question.
These queries are then passed to a specialized translation LLM agent, which converts them into formal graph queries.
This division of labor allows the primary LLM agent to concentrate on high-level reasoning while ensuring syntactically correct and optimized graph queries.
By separating these tasks, \framework enhances query success rates and improves the system's ability to accurately retrieve relevant code information.

\paragraph{Iterative pipeline.}
Instead of completing the code task in a single step, \framework employs an iterative pipeline for interactions between LLM agents and code graph databases, drawing insights from existing agent systems \citep{yao2023react, yang2024swe}.
In each round, LLM agents formulate multiple queries based on the user's question and previously gathered information.
Similar to \cite{madaan2023selfrefine}, the agent then analyzes the aggregated results to determine whether sufficient context has been acquired or if additional rounds are necessary.
This iterative approach fully leverages the reasoning capabilities of the LLM agent, thereby enhancing problem-solving accuracy. \looseness=-1

\section{Experimental Setting}
\paragraph{Benchmarks.}
We employ three diverse repository-level code benchmarks to evaluate \framework: CrossCodeEval \citep{ding2024crosscodeeval}, SWE-bench \citep{yang2024swe}, and EvoCodeBench \citep{li2024evocodebench}.
CrossCodeEval is a multilingual scope cross-file completion dataset for Python, Java, TypeScript, and C\#.
SWE-bench evaluates a model's ability to solve GitHub issues with $2,294$ Issue-Pull Request pairs from $12$ Python repositories.
EvoCodeBench is an evolutionary code generation benchmark with comprehensive annotations and evaluation metrics.

We report our primary results on the CrossCodeEval Lite (Python) and SWE-bench Lite test sets for CrossCodeEval and SWE-bench, respectively, and on the full test set for EvoCodeBench.
CrossCodeEval Lite (Python) and SWE-bench Lite represent subsets of their respective datasets.
CrossCodeEval Lite (Python) consists of $1000$ randomly sampled Python instances, while SWE-bench Lite includes $300$ instances randomly sampled after filtering out those with poor issue descriptions. \looseness=-1

\emph{Remark: During indexing of $43$ Sympy samples from the SWE-bench dataset, we face out-of-memory issues due to numerous files and complex dependencies, leading to their exclusion.
Similarly, some EvoCodeBench samples are omitted due to test environment configuration issues.
Thus, SWE-bench Lite and EvoCodeBench results are based on $257$ and $212$ samples, respectively.} \looseness=-1

\paragraph{Baselines.}
We evaluate whether \framework is a powerful solution for Retrieval-Augmented Code Generation (RACG) \citep{jiang2024survey}.
We specifically assess how effectively code graph database interfaces aid LLMs in understanding code repositories, particularly when handling diverse code questions across different benchmarks to test \framework’s general applicability.
To achieve this, we select resilient RACG baselines that can be adapted to various tasks.
Based on the categories in Section \ref{related_work:RACG}, we choose \textsc{BM25} \citep{stephen09bm25} and \textsc{AutoCodeRover} \citep{zhang2024autocoderover}, which are widely recognized in code tasks \citep{jimenez2023swe, ding2024crosscodeeval, kovrigin2024importance, chen2024coder}, along with a \textsc{No-RAG} method.
Besides, since our work focuses on RACG methods and their generalizability, we exclude methods that interact with external websites \citep{opendevin2024, zhang2024codeagent} and runtime environments \citep{yang2024swe}, as well as task-specific methods that are not easily adaptable across multiple benchmarks \citep{cheng2024dataflow, moatlesstools}.
These methods fall outside the scope of our project. \looseness=-1

Especially, although \cite{zhang2024autocoderover} evaluate \textsc{AutoCodeRover} exclusively on SWE-bench, we extend its implementation to CrossCodeEval and EvoCodeBench, while retaining its core set of $7$ code-specific tools for code retrieval.
\begin{table}[!t]
    \centering
    \caption{\small Performance comparison of \framework and RACG baselines across three benchmarks using different backbone LLMs. The absence of values in SWE-bench Lite for the \textsc{No RAG} method is due to issues with mismatches between the dataset and the code when running inference scripts \protect\footnotemark. Similarly, the missing values in EvoCodeBench are attributable to task inputs being unsuitable for constructing the required \textsc{BM25} queries, and the original paper also does not provide the corresponding implementation. Best results are \textbf{bolded}.}
    \label{main_table}
    \resizebox{1.0\textwidth}{!}{
        \begin{tabular}{llccccccc}
        \toprule
            \multirow{2}*{Model} & \multirow{2}*{Method} & \multicolumn{4}{c}{CrossCodeEval Lite (Python)} & SWE-bench Lite & \multicolumn{2}{c}{EvoCodeBench}\\
            \cmidrule(lr){3-6}\cmidrule(lr){7-7}\cmidrule(lr){8-9}
            ~ & ~ & EM & ES & ID-EM & ID-F1 & Pass@1 & Pass@1 & Recall@1 \\ \toprule
            \multirow{4}*{Qwen2} & \textsc{No RAG} & 8.20 & 46.16 & 13.0 & 36.92 & - & 19.34 & 11.34 \\ 
            ~ & \textsc{BM25} & 15.50 & 51.74 & 22.60 & 45.44 & 0.00 & - & - \\ 
            ~ & \textsc{AutoCodeRover} & 5.21 & 47.63 & 10.16 & 36.54 & 9.34 & 16.91 & 7.86 \\
            % \rowcolor{gray!15}
            \cmidrule(lr){2-2}
            ~ & \framework & 5.00 & 47.99 & 9.10 & 36.44 & 1.95 & 14.62 & 8.60 \\ 
            \midrule
            \multirow{4}*{DS-Coder} & \textsc{No RAG} & 11.70 & 60.73 & 16.90 & 47.85 & - & 25.47 & 11.04 \\ 
            ~ & \textsc{BM25} & 21.90 & 67.52 & 30.60 & 59.04 & 1.17 & - & - \\ 
            ~ & \textsc{AutoCodeRover} & 14.90 & 59.78 & 22.30 & 51.34 & 15.56 & 20.28 & 7.56 \\ 
            % \rowcolor{gray!15}
            \cmidrule(lr){2-2}
            ~ & \framework & 20.20 & 63.14 & 28.10 & 54.88 & 12.06 & 27.62 & \textbf{12.01} \\ 
            \midrule
            \multirow{4}*{GPT-4o} & \textsc{No RAG} & 10.80 & 59.36 & 16.70 & 48.22 & - & 27.83 & 11.79 \\ 
            ~ & \textsc{BM25} & 21.20 & 66.18 & 30.20 & 58.71 & 3.11 & - & - \\ 
            ~ & \textsc{AutoCodeRover} & 21.20 & 61.92 & 28.10 & 54.81 & \textbf{22.96} & 28.78 & 11.17 \\
            % \rowcolor{gray!15}
            \cmidrule(lr){2-2}
            ~ & \framework & \textbf{27.90} & \textbf{67.98} & \textbf{35.60} & \textbf{61.08} & \textbf{22.96} & \textbf{36.02} & 11.87 \\ 
        \bottomrule
        \vspace{-2em}
        \end{tabular}
    }
\end{table}
\footnotetext{\url{https://github.com/princeton-nlp/SWE-bench/issues/2}}

\paragraph{Large Language Models (LLMs).}
We evaluate \framework on three advanced and well-known LLMs with long text processing, tool use, and code generation capabilities: GPT-4o, DeepSeek-Coder-V2 \citep{zhu2024deepseek}, and Qwen2-72b-Instruct \citep{yang2024qwen2}.
\begin{itemize}[leftmargin=12pt, nosep]
    \item \textbf{GPT-4o}: Developed by OpenAI \footnote{We use the \texttt{gpt-4o-2024-05-13} version, \url{https://openai.com/api}}, this model excels in commonsense reasoning, mathematics, and code, and is among the top-performing models as of July 2024 \footnote{\url{https://huggingface.co/spaces/lmsys/chatbot-arena-leaderboard}}.
    \item \textbf{DeepSeek-Coder-V2 (DS-Coder)}: A specialized code-specific LLM by DeepSeek \footnote{\url{https://chat.deepseek.com/coder}}, it retains general capabilities while being highly proficient in code-related tasks.
    \item \textbf{Qwen2-72b-Instruct (Qwen2)}: Developed by Alibaba \footnote{\url{https://dashscope.console.aliyun.com/model}}, this open-source model has about $72$ billion parameters and a 128k long context, making it suitable for evaluating existing methods.

\end{itemize}

For the hyperparameters of the selected large language models, we empirically set the temperature coefficient to $0.0$ for both GPT-4o and Qwen2-72b-Instruct, and to $1.0$ for DeepSeek-Coder-V2.
All other parameters are kept at their default settings.

\paragraph{Metrics.}
In metrics selection, we follow the original papers' settings \citep{jimenez2023swe, ding2024crosscodeeval, li2024evocodebench}.
Specifically, for CrossCodeEval, we measure performance with code match and identifier match metrics, assessing accuracy with exact match (EM), edit similarity (ES), and F1 scores.
SWE-bench utilizes \% Resolved (Pass@$1$) to gauge the effectiveness of model-generated patches based on provided unit tests.
EvoCodeBench employs Pass@$k$, where $k$ represents the number of generated programs, for functional correctness and Recall@$k$ to assess the recall of reference dependencies in generated programs.
We set $k$ to $1$ in our main experiments.

\paragraph{Implementation details.}
Before indexing, we filter the Python repositories for each benchmark to retain only Python files.
For the SWE-bench dataset, we also exclude test files to avoid slowing down the creation of the code graph database.
Following the process outlined in Section \ref{build_code_graph_db}, we construct code graph databases for the indexed repositories, storing the corresponding nodes and edges.
We select Neo4j as the graph database and Cypher as the query language.

\section{Results}
\label{exp:results}
\subsection{Analysis of Repository-Level Code Tasks}
\textbf{RACG is crucial for repository-level code tasks.} In Table \ref{main_table}, RACG-based methods—\textsc{BM25}, \textsc{AutoCodeRover}, and \textsc{CodexGraph}—basically outperform the \textsc{No-RAG} method across all benchmarks and evaluation metrics.
For instance, on the CrossCodeEval Lite (Python) dataset, using GPT-4o as the backbone LLM, RACG methods improve performance by 10.4\% to 17.1\% on the EM metric compared to \textsc{No-RAG}.
This demonstrates that the \textsc{No-RAG} approach, which relies solely on in-file context and lacks interaction with the code repository, significantly limits performance.

\textbf{Existing RACG methods struggle to adapt to various repo-level code tasks.} Experimental results in Table \ref{main_table} reveal the shortcomings of existing RACG-based methods like \textsc{BM25} and \textsc{AutoCodeRover}.
While these methods perform well in specific tasks, they often underperform when applied to other repository-level code tasks.
This discrepancy typically arises from their inherent characteristics or task-specific optimizations.

Specifically, \textsc{AutoCodeRover} is designed with code tools tailored for SWE-bench tasks, leveraging expert knowledge and the unique features of SWE-bench to optimize tool selection and design.
This optimization refines the LLM agent's action spaces, enabling it to gather valuable information more efficiently and boosting its performance on SWE-bench tasks ($22.96$\%).
However, these task-specific optimizations limit its flexibility and effectiveness in other coding tasks, as evidenced by its subpar results on CrossCodeEval Lite (Python) and EvoCodeBench compared to other methods.

Similarly, \textsc{BM25} faces the same issues.
In CrossCodeEval Lite (Python), its similarity-based retrieval aligns well with code completion tasks, enabling it to easily retrieve relevant usage references or direct answers.
This results in strong performance, particularly in the ES metric.
However, \textsc{BM25} lacks the reasoning capabilities of LLMs during query construction, making its retrieval process less intelligent.
Consequently, when confronted with reasoning-heavy tasks like those in SWE-bench, \textsc{BM25} often fails to retrieve appropriate code snippets, leading to poor performance.

\textbf{\framework shows versatility and efficacy across diverse benchmarks.} Table \ref{main_table} shows that \framework achieves competitive results across various repository-level code tasks with general code graph database interfaces.
Specifically, with GPT-4o as the LLM backbone, \framework outperforms other RACG baselines on CrossCodeEval Lite (Python) and EvoCodeBench, while also achieving results comparable to \textsc{AutoCodeRover} on SWE-bench Lite.
This demonstrates the generality and effectiveness of the code graph database interface design.

\begin{table}[!t]
    \centering
        \caption{\small Average token cost comparison across three benchmarks (GPT-4o as the backbone LLM).}
        \begin{tabular}{cccc}
        \toprule
        ~ & CrossCodeEval Lite (Python) & SWE-bench Lite & EvoCodeBench \\ 
        \midrule
        \textsc{BM25} & 1.47k& 14.76k& -\\ 
        \textsc{AutoCodeRover} & 10.74k& 76.01k& 21.41k\\ 
        \framework & 22.16k& 102.25k& 24.49k\\
        \bottomrule
        \end{tabular}
    \label{tab:token_cost}
\end{table}

\textbf{\framework increases token consumption.} \framework uses code graph databases as interfaces and retrieves information from the code repository by writing graph queries.
While benefiting from larger and more flexible action spaces, it also incurs increased token costs.
The primary reason for this is that the length of the query outcomes is not controllable.
Moreover, \framework sometimes encounters loops where it fails to generate executable graph queries.
As demonstrated in Table \ref{tab:token_cost}, this leads to a higher token usage compared to existing RACG methods. \looseness=-1

\subsection{Deeper Analysis of \framework}
\begin{wrapfigure}{r}{0.5\textwidth}
    \centering
    \vspace{-1em}
    \includegraphics[width=0.45\textwidth]{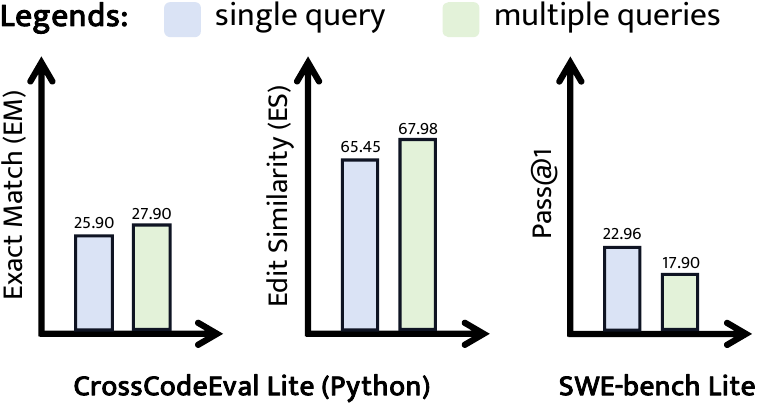}
    \caption{\small Performance comparison of different querying strategies on CrossCodeEval Lite (Python) and SWE-bench Lite.}
    \vspace{-1em}
    \label{fig:query_num}
\end{wrapfigure}
\textbf{Optimal querying strategies vary across different benchmarks.} There are two strategies for formulating queries in each round within \framework: either generating a single query or producing multiple queries for code retrieval.
Opting for a single query per round can enhance precision in retrieving relevant content but may compromise the recall rate.
Conversely, generating multiple queries per round can improve recall but may reduce precision.
Experimental results, as illustrated in Figure \ref{fig:query_num}, reveal that for CrossCodeEval Lite (Python), which involves lower reasoning difficulty ($26.43$ \textit{vs.} $27.90$ in the EM metric), the ``multiple queries'' strategy is more effective.
In contrast, for SWE-bench Lite, which presents higher reasoning difficulty, the ``single query'' strategy yields better outcomes ($22.96$ \textit{vs.} $17.90$ in the Pass@$1$ metric).
These findings provide valuable guidance for researchers in selecting the most appropriate querying strategy for future studies.

\begin{table}[!t]
    \centering
        \caption{\small Ablation study about the translation LLM agent on CrossCodeEval Lite (Python).}
        \vspace{0.5em}
        \resizebox{1.0\textwidth}{!}{
            \begin{tabular}{llcccc}
            \toprule
            \multirow{2}*{Model} & \multirow{2}*{Method} & \multicolumn{4}{c}{CrossCodeEval Lite (Python)} \\
            \cmidrule(lr){3-6}
            ~ & ~ & EM & ES & ID-EM & ID-F1 \\
            \midrule
            \multirow{2}*{Qwen2} & \framework & 5.00 & 47.99 & 9.10 & 36.44 \\
            ~ & \textit{w/o translation LLM Agent} & 0.50 \textcolor{deepgreen}{(-4.50)} & 10.45 \textcolor{deepgreen}{(-37.54)} & 0.60 \textcolor{deepgreen}{(-8.50)} & 2.62 \textcolor{deepgreen}{(-33.82)}\\
            \midrule
            \multirow{2}*{DS-Coder} & \framework & 20.20 & 63.14 & 28.10 & 54.88 \\
            ~ & \textit{w/o translation LLM Agent} & 5.50 \textcolor{deepgreen}{(-14.70)} & 53.56 \textcolor{deepgreen}{(-9.58)} & 11.20 \textcolor{deepgreen}{(-16.90)} & 39.75 \textcolor{deepgreen}{(-15.13)}\\
            \midrule
            \multirow{2}*{GPT-4o} & \framework & 27.90 & 67.98 & 35.60 & 61.08 \\
            ~ & \textit{w/o translation LLM Agent} & 8.30 \textcolor{deepgreen}{(-19.60)} & 56.36 \textcolor{deepgreen}{(-11.62)} & 14.40 \textcolor{deepgreen}{(-21.20)} & 44.08 \textcolor{deepgreen}{(-17.00)}\\
            \bottomrule
            \end{tabular}
        }
    \label{tab:cypher_agent}
\end{table}

\textbf{``Write then translate'' eases reasoning load.} When the assistance of the translation LLM agent is removed, the primary LLM agent must independently analyze the coding question and directly formulate the graph query for code retrieval.
This increases the reasoning load on the primary LLM agent, leading to a decline in the syntactic accuracy of the graph queries.
Experimental results in Table \ref{tab:cypher_agent} highlight the significant negative impact of the removal of the translation LLM agent on \framework's performance across all selected LLMs in the CrossCodeEval Lite (Python) benchmark.
Even when GPT-4o is used as the backbone model, performance metrics exhibit a significant drop (e.g., the EM metric drops from 27.90\% to 8.30\%), underscoring the critical role of the translation LLM agent in alleviating the primary LLM agent's reasoning burden.

\textbf{\framework is enhanced when equipped with advanced LLMs.}
Code graph databases provide a flexible and general interface, resulting in a broader action space for \framework compared to existing methods.
However, if the underlying LLM lacks sufficient reasoning and coding capabilities, the LLM agent in \framework may struggle to formulate appropriate graph queries.
This can lead to failures in retrieving the expected code, which in turn hampers further reasoning.

As shown in Table \ref{main_table}, the effectiveness of \framework improves significantly with advancements in LLMs.
For example, transitioning from Qwen2-72b-Instruct to DeepSeek-Coder-v2 and then to GPT-4o, the overall performance enhancement across various benchmarks and metrics is notable.
This illustrates that while \framework requires high-level coding skills, reasoning abilities, and proficiency in handling complex texts from LLMs, the rapid advancement of these models allows them to better leverage the flexible interfaces provided by code graph databases.

\section{Real-World Application Scenario}
\label{sec:application}
To highlight the practical value of the \framework in real-world applications, we develop five code agents using the flexible ModelScope-Agent framework \citep{li2023modelscopeagent}.
These agents are designed to address common coding challenges in production environments by integrating key concepts of the \framework.
\textbf{Code Chat} allows users to inquire about a code repository, providing insights into code structure and function usage.
\textbf{Code Debugger} diagnoses and resolves bugs by applying iterative reasoning and information retrieval to suggest targeted fixes.
\textbf{Code Unittestor} generates unit tests for specified classes or functions to ensure thorough functionality verification.
\textbf{Code Generator} automatically creates code to meet new requirements, extending the functionality of existing codebases.
Lastly, \textbf{Code Commentor} produces comprehensive annotations, enhancing documentation for code segments lacking comments.
Examples of these agents are provided in Appendix \ref{app:code_agent} to maintain brevity in the main text.

\section{Discussion}
\paragraph{Limitations.} \framework has only been evaluated on a single programming language, Python.
In the future, we plan to extend \framework to more programming languages, such as Java and C++.
Secondly, there is room for improvement in the construction efficiency and schema completeness of the code graph database.
Faster database indexing and a more comprehensive schema (e.g., adding edges related to function calls) will enhance the broader applicability of \framework.
Finally, the design of \framework's workflow can further integrate with existing advanced agent techniques, such as finer-grained multi-agent collaboration.

\paragraph{Conclusion.} \framework addresses the limitations of existing RACG methods, which often struggle with flexibility and generalization across different code tasks.
By integrating LLMs with code graph database interfaces, \framework facilitates effective, code structure-aware retrieval for diverse repository-level code tasks.
Our evaluations highlight its competitive performance and broad applicability on academic benchmarks.
Additionally, we provide several code applications in ModelScope-Agent, demonstrating \framework’s capability to enhance the accuracy and usability of automated software development.

\newpage
\bibliography{neurips_2024}
\bibliographystyle{neurips_2024}

\newpage

\appendix
\section{Appendix}
\subsection{Details of the graph database schema}
\label{app:schema}
This schema is designed to abstract code repositories into code graphs for Python, where nodes represent symbols in the source code, and edges represent relationships between these symbols.

\subsubsection{Node Types}

Each node in the code graph represents a different element within Python code, and each node type has a set of attributes that encapsulate its meta-information. The node types and their respective attributes are as follows:
\begin{codebox}[Graph Database Schema: Nodes]
\begin{lstlisting}[language=]
## Nodes
MODULE:
  Attributes:
    - name (String): Name of the module (dotted name)
    - file_path (String): File path of the module

CLASS:
  Attributes:
    - name (String): Name of the class
    - file_path (String): File path of the class
    - signature (String): The signature of the class
    - code (String): Full code of the class

FUNCTION:
  Attributes:
    - name (String): Name of the function
    - file_path (String): File path of the function
    - code (String): Full code of the function
    - signature (String): The signature of the function

FIELD:
  Attributes:
    - name (String): Name of the field
    - file_path (String): File path of the field
    - class (String): Name of the class the field belongs to

METHOD:
  Attributes:
    - name (String): Name of the method
    - file_path (String): File path of the method
    - class (String): Name of the class the method belongs to
    - code (String): Full code of the method
    - signature (String): The signature of the method

GLOBAL_VARIABLE:
  Attributes:
    - name (String): Name of the global variable
    - file_path (String): File path of the global variable
    - code (String): The code segment in which the global variable is defined
\end{lstlisting}
\end{codebox}

\newpage
\subsubsection{Edge Types}
Edges in the code graph represent various relationships between the nodes. The edge types we define and the relationships they signify are as follows:
\begin{codebox}[Graph Database Schema: Edges]
\begin{lstlisting}[language=]
## Edges
CONTAINS:
  Source: MODULE
  Target: CLASS or FUNCTION or GLOBAL_VARIABLE

HAS_METHOD:
  Source: CLASS
  Target: METHOD

HAS_FIELD:
  Source: CLASS
  Target: FIELD

INHERITS:
  Source: CLASS
  Target: CLASS (base class)

USES:
  Source: FUNCTION or METHOD
  Target: GLOBAL_VARIABLE or FIELD
  Attributes:
    - source_association_type (String): FUNCTION, METHOD
    - target_association_type (String): GLOBAL_VARIABLE, FIELD
\end{lstlisting}
\end{codebox}

\newpage
\subsection{Real-World Application}
\label{app:code_agent}

In this section, we present the WebUI interface for \framework, showcasing its five practical applications: Code Chat, Code Debugger, Code Unittestor, Code Generator, and Code Commentor.
The interface is designed to facilitate user interaction, providing a streamlined and intuitive environment for various code-related tasks.
We built the WebUI interface using Streamlit\footnote{Streamlit: \url{https://streamlit.io/}}, a powerful and user-friendly framework that allows for the rapid development of interactive web applications.

\begin{figure}[h]
    \centering
    \includegraphics[width=0.9\textwidth]{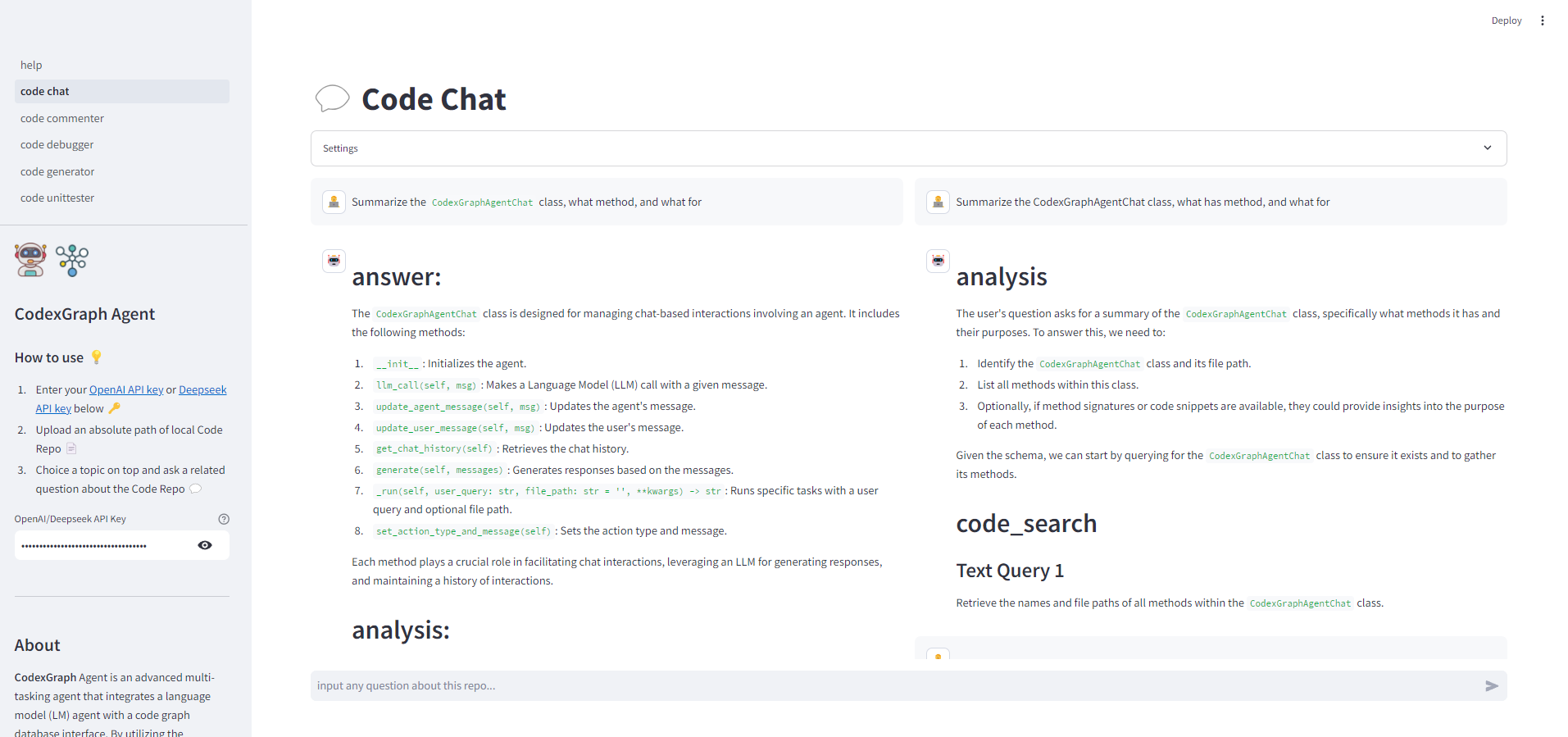}
    \caption{WebUI for the Code Chat, used for answering any questions related to code repositories.}
    \label{fig:code_chat}
\end{figure}

\begin{figure}[h]
    \centering
    \subcaptionbox{\footnotesize Code Debugger}{
        \includegraphics[width=0.46\linewidth]{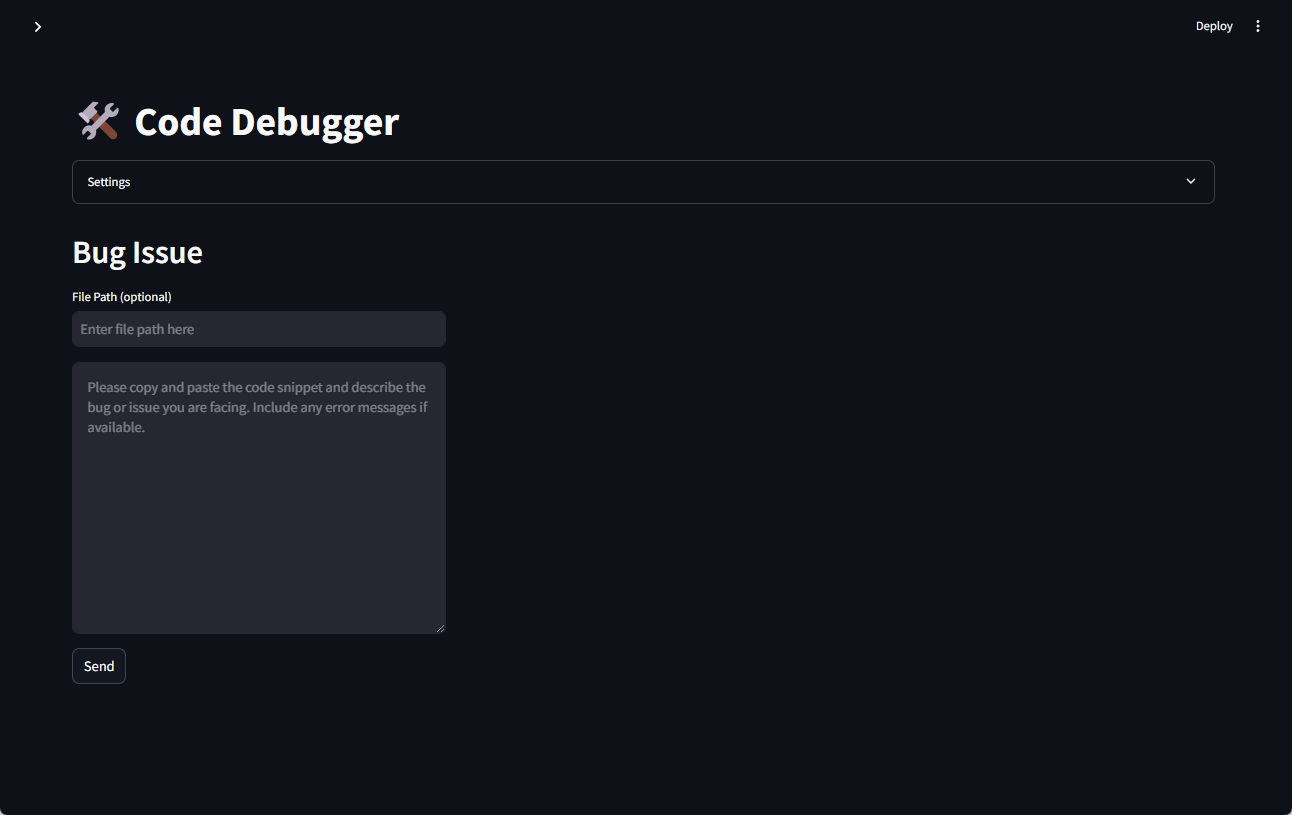} 
        % \hspace{-6mm}
        % \vspace{-1mm}
    }
    \subcaptionbox{\footnotesize Code Unittestor}{
        \includegraphics[width=0.46\linewidth]{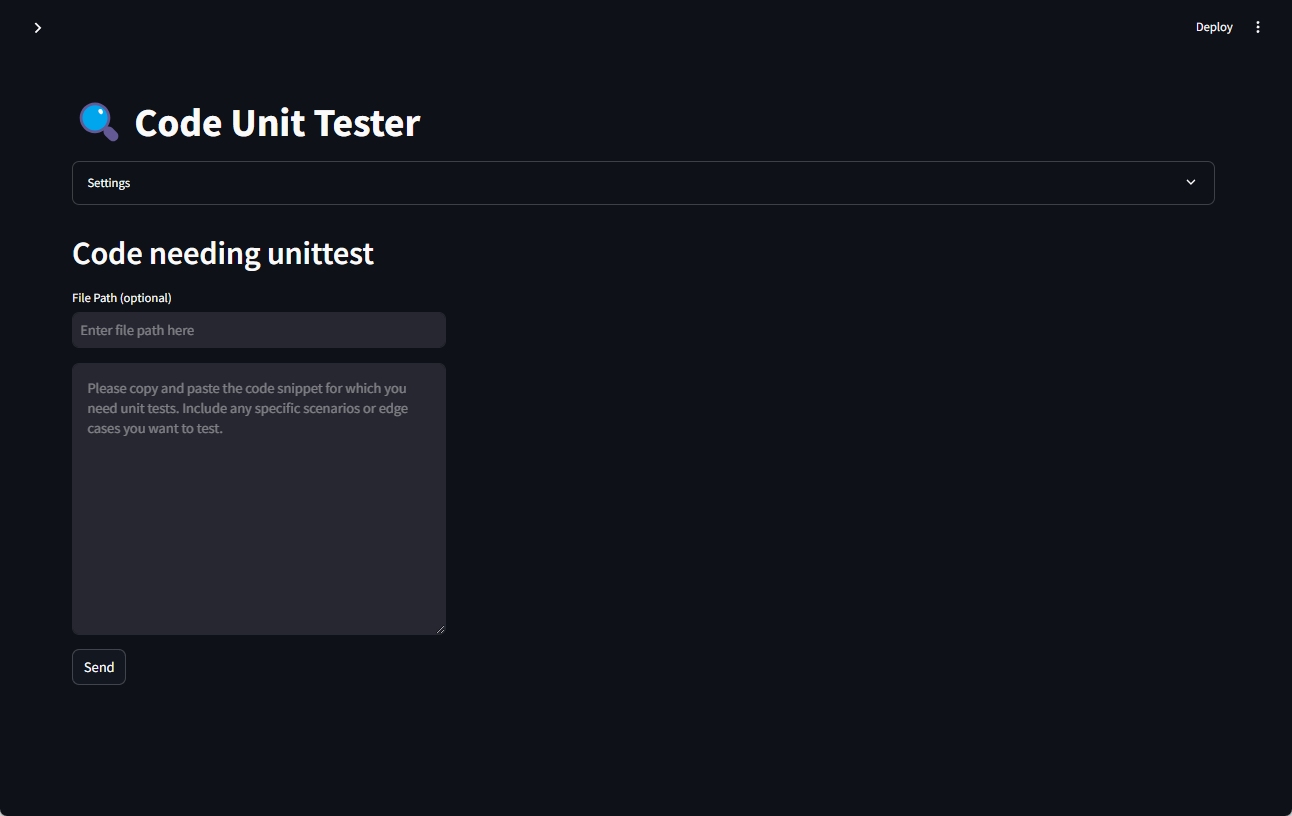} 
        % \hspace{-6mm}
        % \vspace{-1mm}
    }
    \\
    \subcaptionbox{\footnotesize Code Generator}{
        \includegraphics[width=0.46\linewidth]{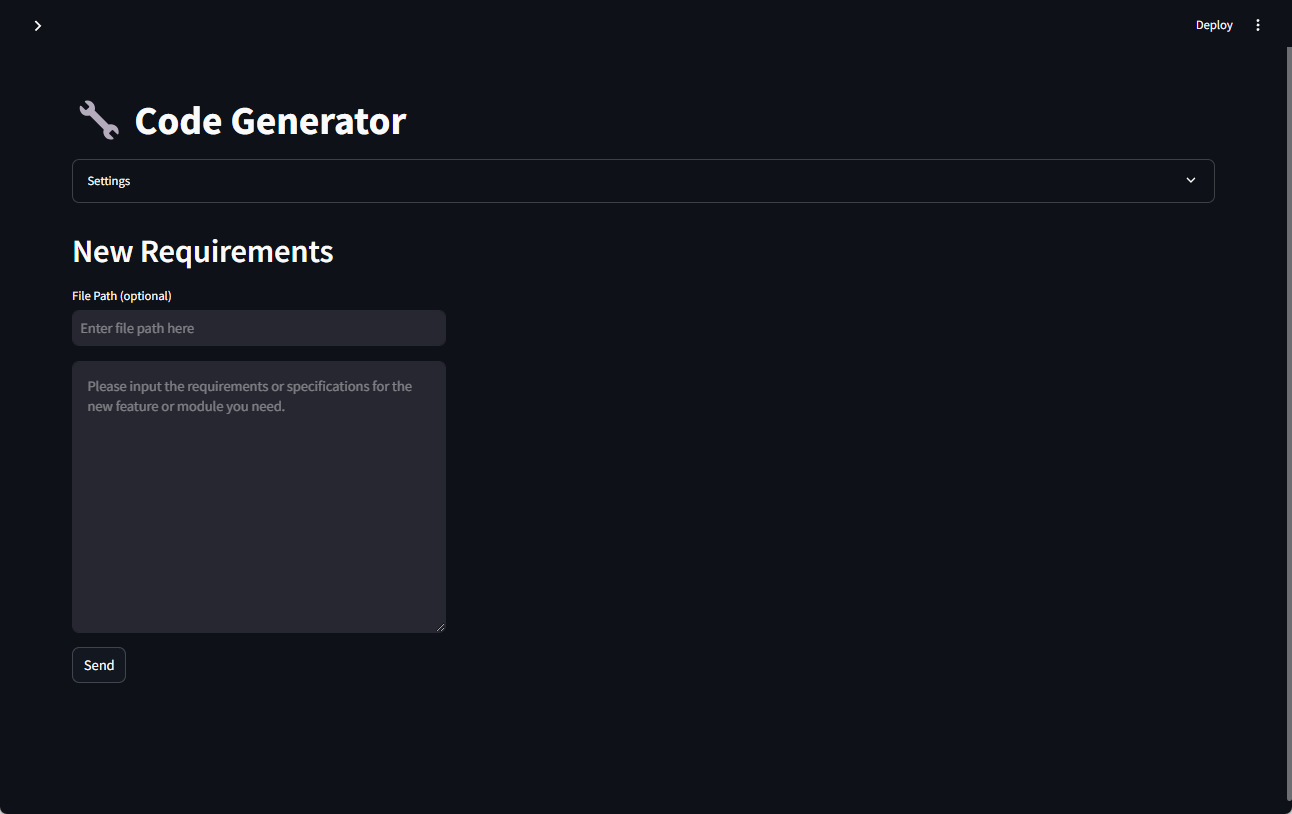} 
        % \hspace{-6mm}
        % \vspace{-1mm}
    }
    \subcaptionbox{\footnotesize Code Commentor}{
        \includegraphics[width=0.46\linewidth]{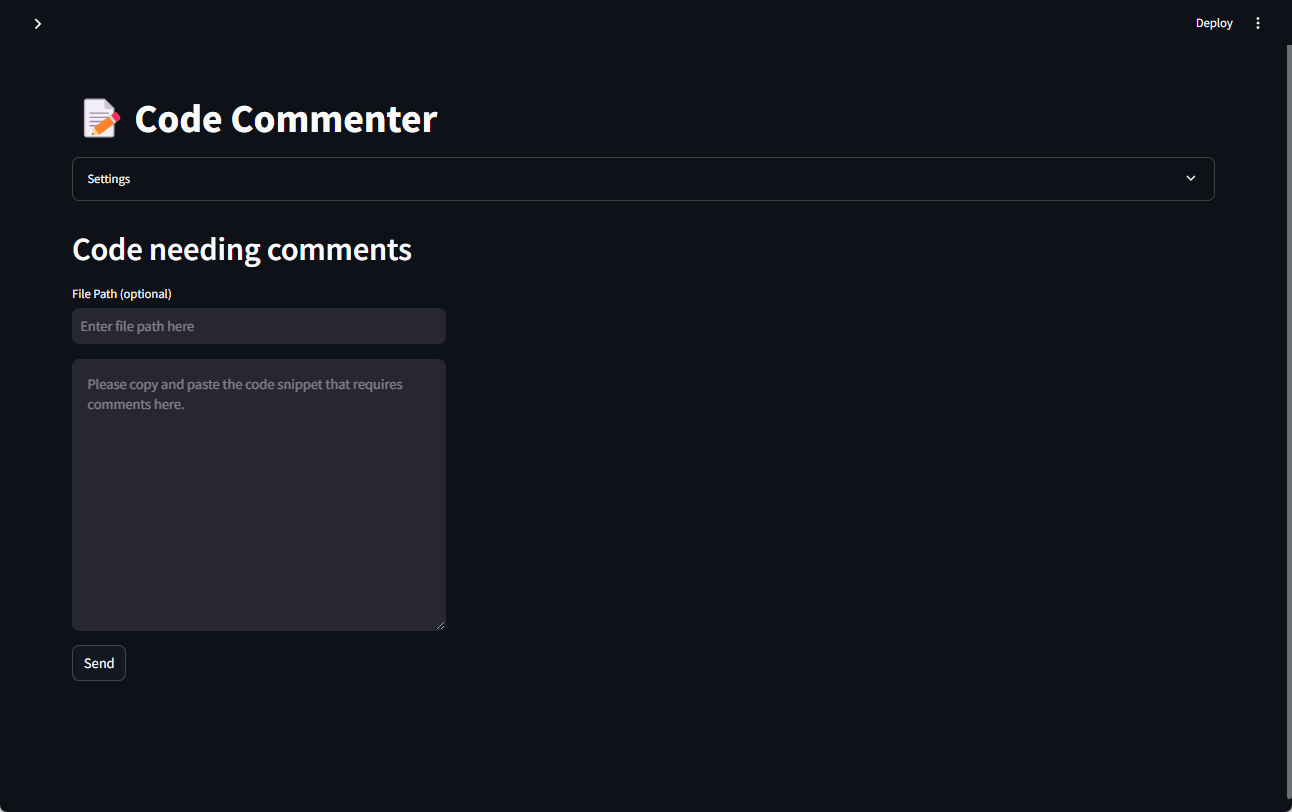} 
        % \hspace{-6mm}
        % \vspace{-1mm}
    }
    \caption{WebUI for Code Debugger, Code Unittestor, Code Generator, and Code Commentor.}
    \label{fig: intro}
\end{figure}

To experience our application firsthand, you can visit ModelScope-Agent and navigate to the \framework \footnote{ \url{https://github.com/modelscope/modelscope-agent/tree/master/apps/codexgraph_agent}}.
This repository provides a detailed guide on how to set up and interact with the various applications we have described.

\newpage
\subsubsection{Example of Code Chat}
Code Chat allows users to inquire about a code repository, providing insights into code structure and function usage. This functionality is particularly useful for understanding complex codebases, identifying dependencies, and exploring the usage of specific classes, methods, and functions.

Here is an example of Code Chat. The user's question is ``
Summarize the `CodexGraphAgentChat' class, what has method, and what for''.

\begin{figure}[h]
    \centering
    \includegraphics[width=0.7\textwidth]{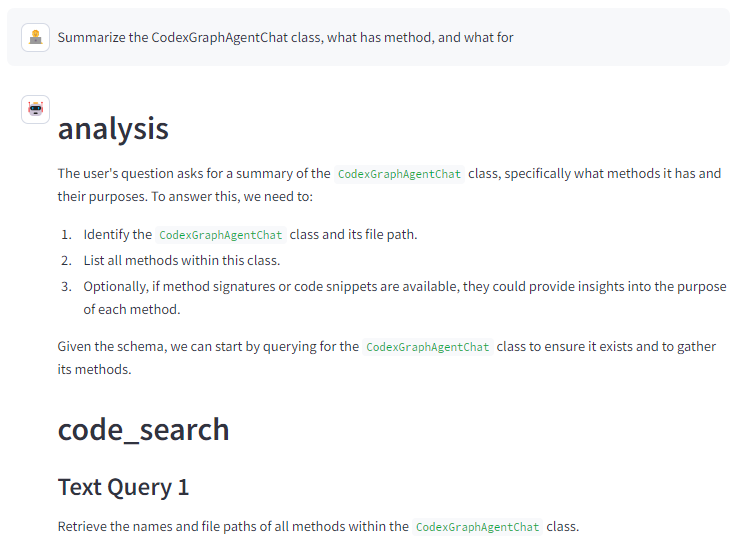}
    \caption{Using Cypher queries to retrieve information about the `CodexGraphAgentChat' class, from the code repository.}
    \label{fig:code_chat1}
\end{figure}

\begin{figure}[h]
    \centering
    \includegraphics[width=0.7\textwidth]{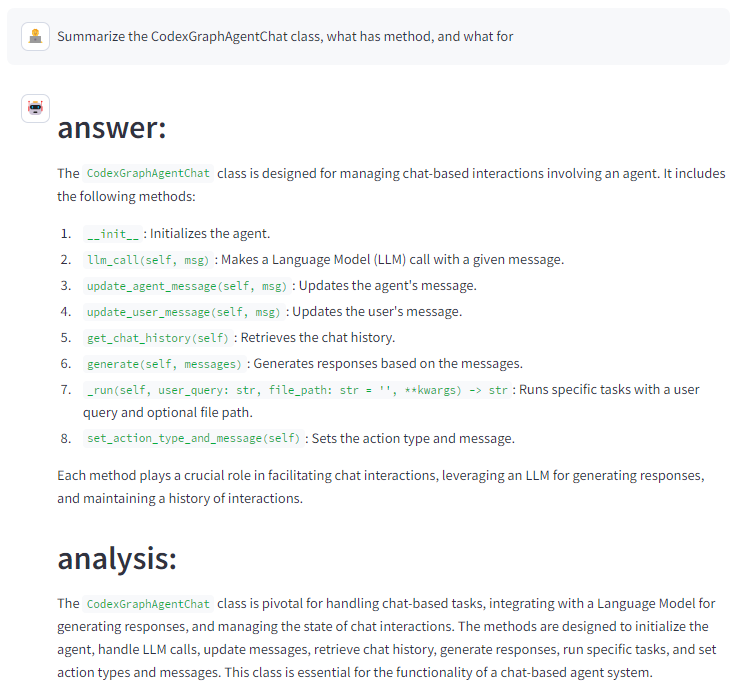}
    \caption{Once the necessary information is gathered, Code Chat constructs a comprehensive response to the user's question. This response includes a summary of the `CodexGraphAgentChat' class, a list of its methods, and a description of what each method does.}
    \label{fig:code_chat2}
\end{figure}

\newpage
\subsubsection{Example of Code Debugger}

The Code Debugger diagnoses and resolves bugs by applying iterative reasoning and information retrieval to suggest targeted fixes. It utilizes Cypher queries to analyze the code repository, identify the cause of the issue, and recommend precise modifications.

Here is an example of Code Debugger. The user's input is a real issue\footnote{\url{https://github.com/modelscope/modelscope-agent/pull/549}} where the outcome does not match the expected behavior. The Code Debugger first analyzes the problem, then uses Cypher queries to retrieve relevant information and infer the cause of the bug. Finally, it provides an explanation of the bug and suggests the location for the modification.

\begin{figure}[h]
    \centering
    \includegraphics[width=0.85\textwidth]{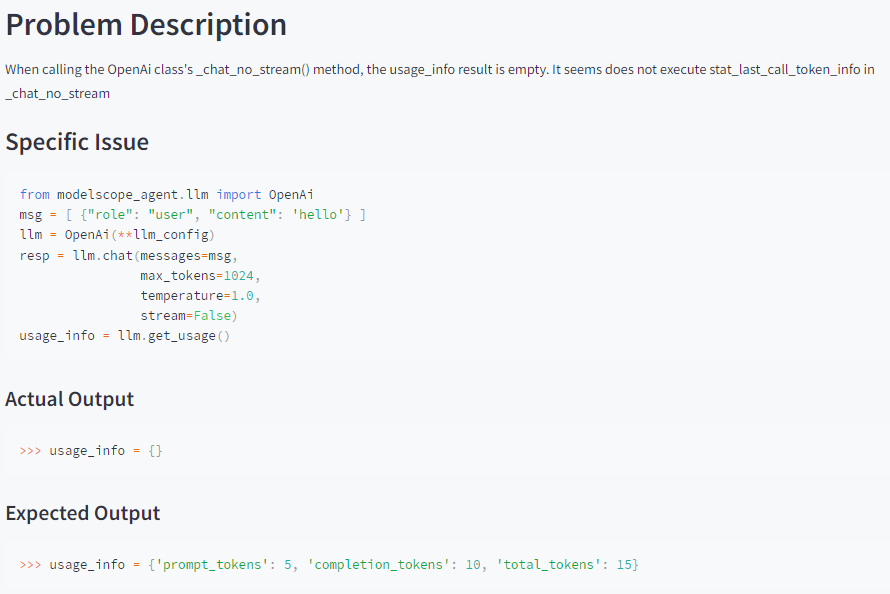}
    \caption{The issue describes a problem where the outcome does not match the expected behavior.}
    \label{fig:code_debug1}
\end{figure}

\begin{figure}[h]
    \centering
    \includegraphics[width=0.9\textwidth]{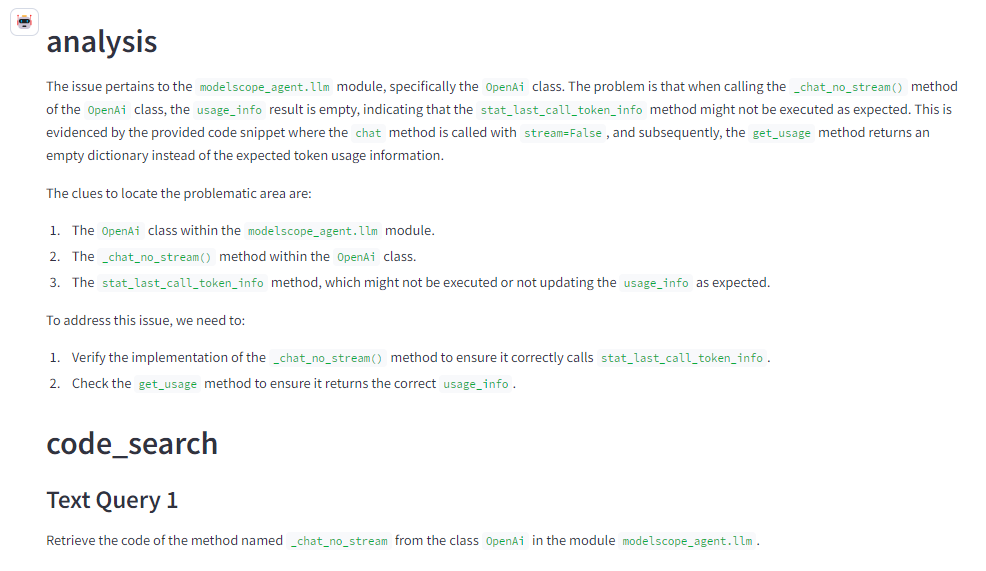}
    \caption{Analyzing the problem and retrieving information using Cypher queries.}
    \label{fig:code_debug2}
\end{figure}

\begin{figure}[t]
    \centering
    \includegraphics[width=0.9\textwidth]{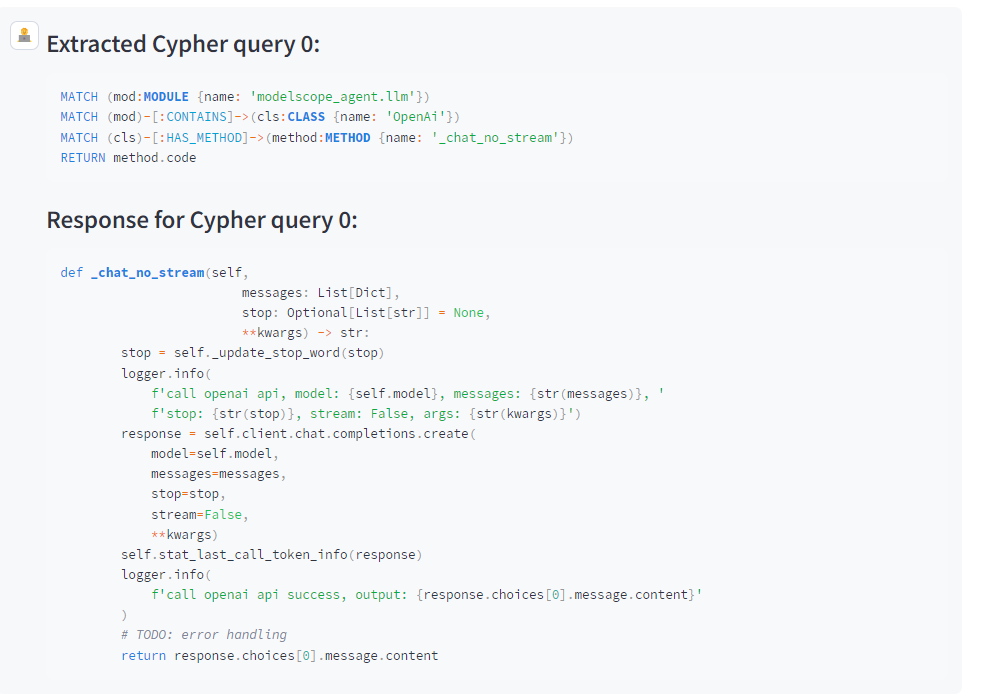}
    \caption{Executing Cypher queries to search the code for relevant information.}
    \label{fig:code_debug3}
\end{figure}

\begin{figure}[t]
    \centering
    \includegraphics[width=0.9\textwidth]{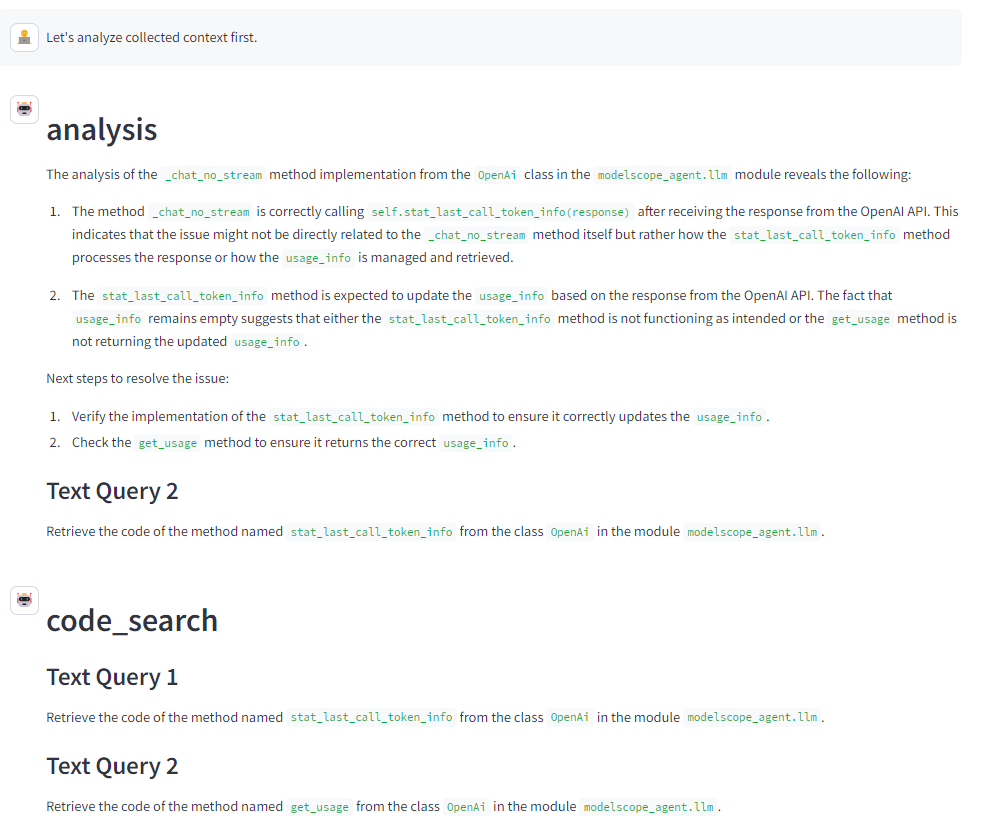}
    \caption{Analyzing the retrieved information to identify potential causes of the bug.}
    \label{fig:code_debug4}
\end{figure}

\begin{figure}[t]
    \centering
    \includegraphics[width=0.9\textwidth]{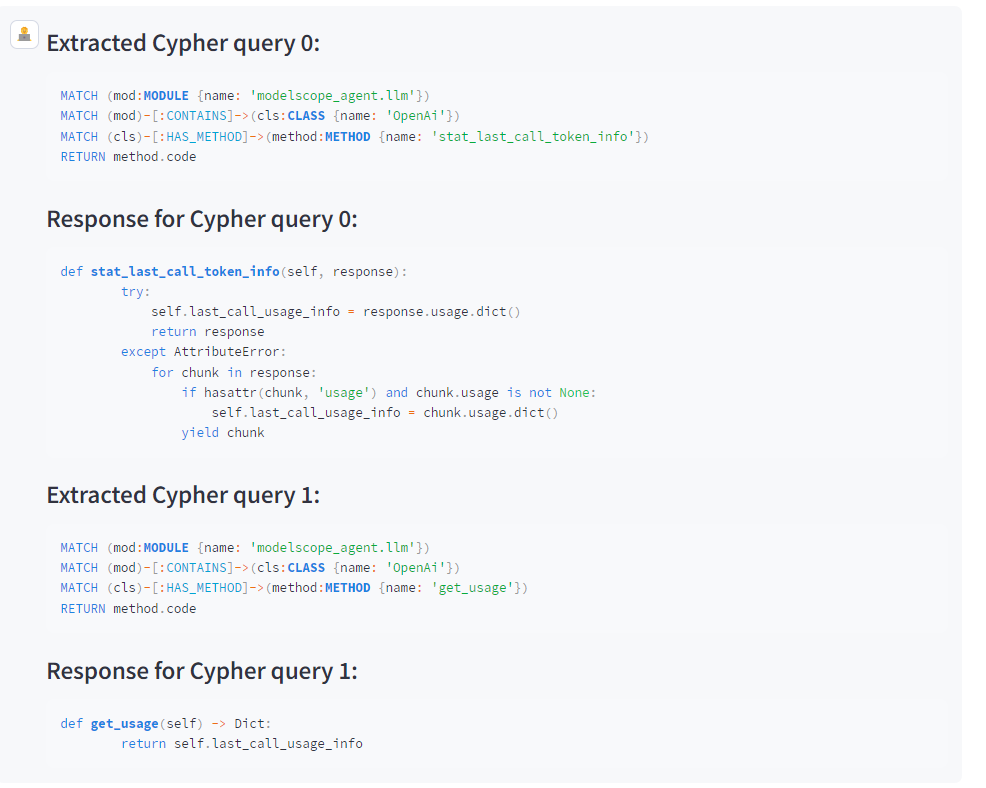}
    \caption{Performing additional Cypher code searches to gather more information.}
    \label{fig:code_debug5}
\end{figure}

\begin{figure}[t]
    \centering

    \includegraphics[width=0.9\textwidth]{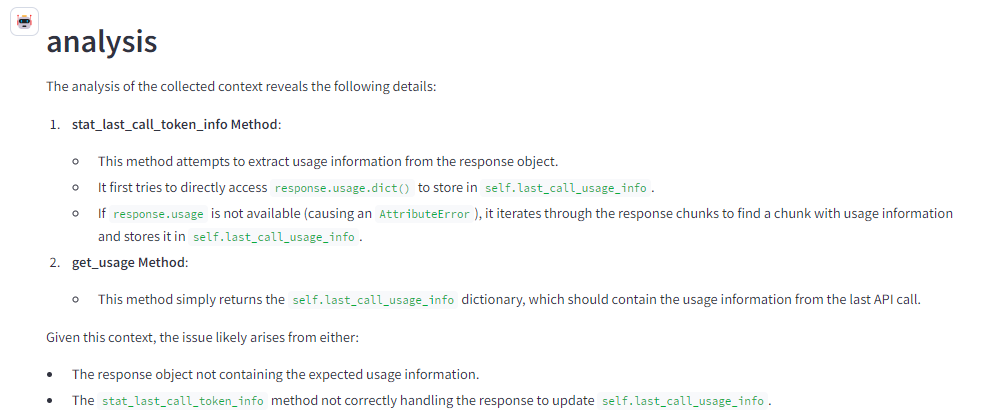}
    \caption{Inferring the cause of the bug based on the analysis of the retrieved information.}
    \label{fig:code_debug6}
\end{figure}

\begin{figure}[t]
    \centering
    \includegraphics[width=0.9\textwidth]{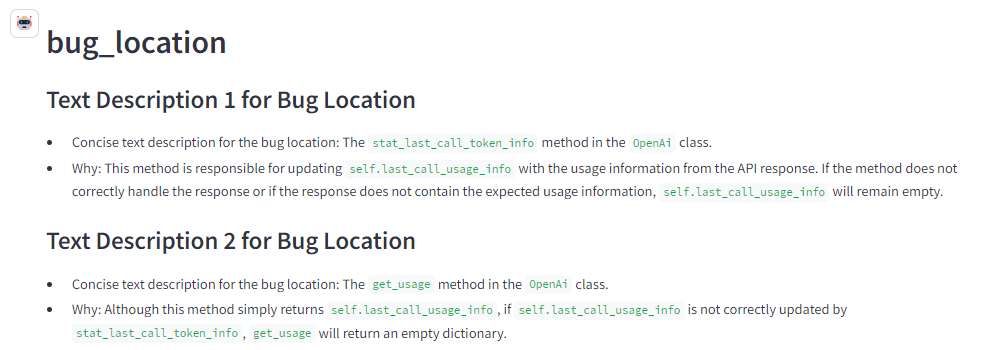}
    \caption{Identifying the precise location of the bug in the codebase.}
    \label{fig:code_debug7}
\end{figure}

\begin{figure}[t]
    \centering
    \includegraphics[width=0.9\textwidth]{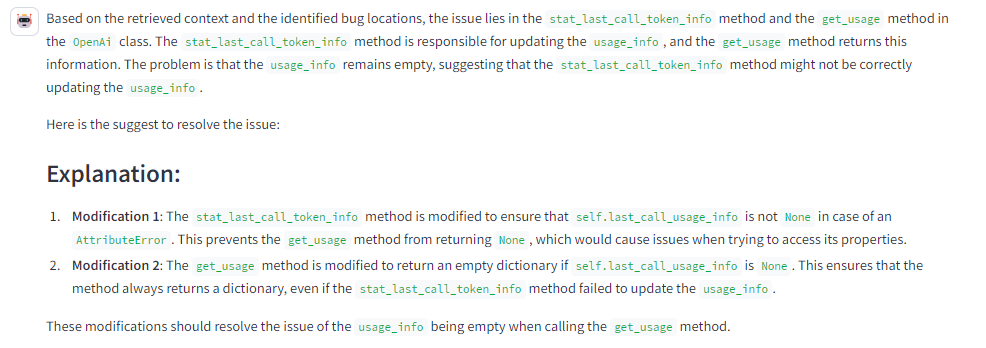}
    \caption{Providing a detailed explanation of the issue and the underlying cause of the bug.}
    \label{fig:code_debug8}
\end{figure}

\begin{figure}[t]
    \centering
    \includegraphics[width=0.9\textwidth]{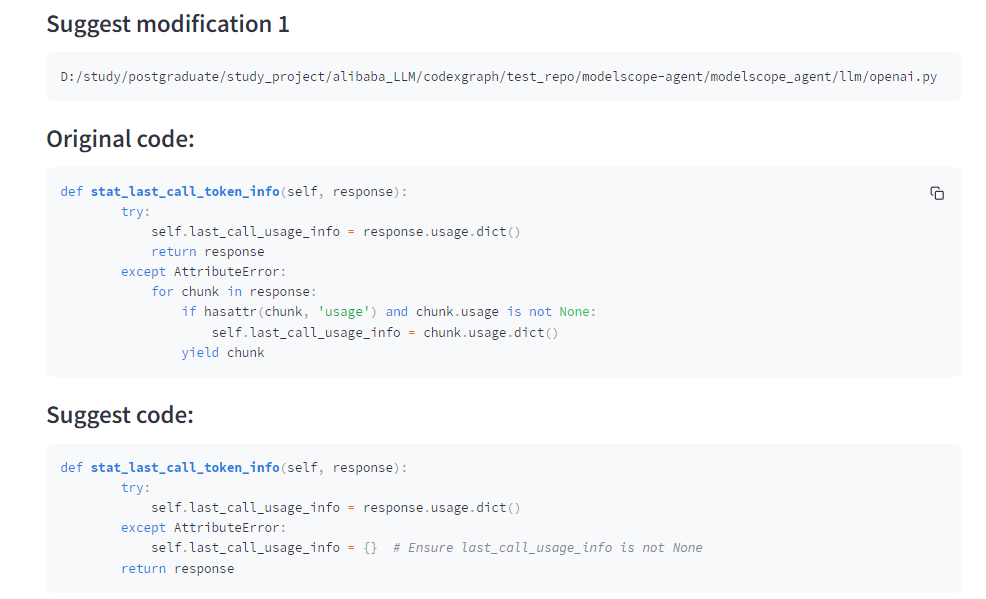}
    \caption{Suggesting the first modification to resolve the bug.}
    \label{fig:code_debug9}
\end{figure}

\begin{figure}[t]
    \centering
    \includegraphics[width=0.9\textwidth]{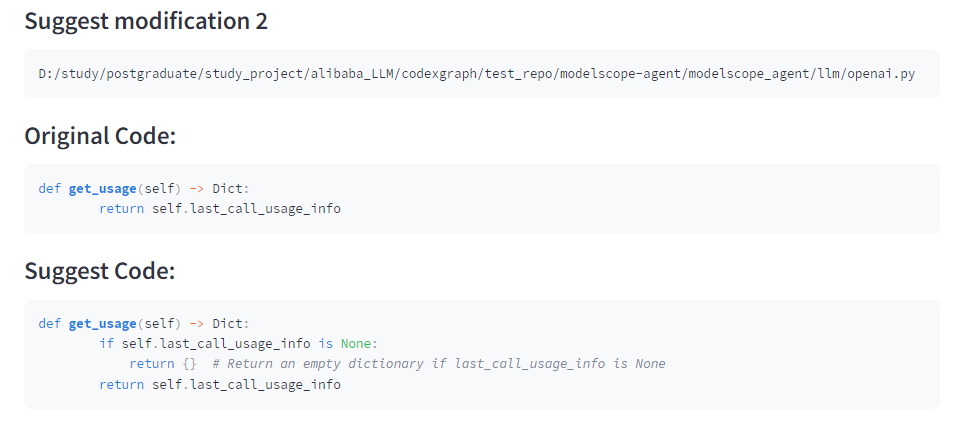}
    \caption{Suggesting the second modification to ensure the bug is resolved.}
    \label{fig:code_debug10}
\end{figure}

% In this example, the Code Debugger systematically diagnoses the problem, retrieves pertinent information, and suggests targeted fixes, helping developers resolve issues efficiently and effectively.

\clearpage
\subsubsection{Example of Code Unittestor}

Here is an example of Code Unittestor. The user's input is: "Generate test cases for TaskManager." The CodexGraph agent will first retrieve all methods and inheritance relationships in `TaskManager', and then generate detailed test case code.

\begin{figure}[h]
    \centering
    \includegraphics[width=0.85\textwidth]{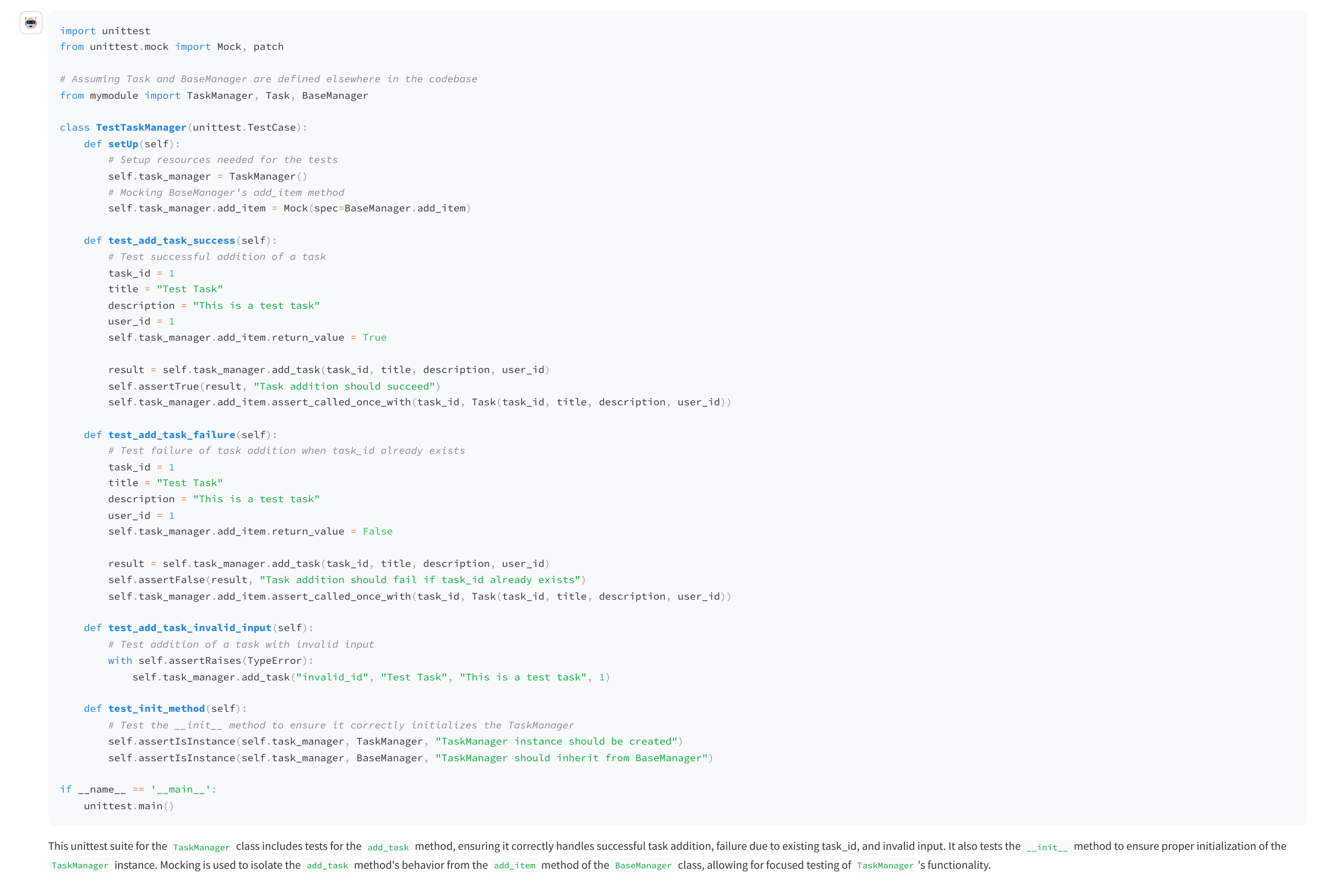}
    \caption{Generated detailed unit test code for the `TaskManager' class, covering its methods and inheritance relationships.}
\end{figure}

\subsubsection{Example of Code Generator}
The user has requested a function to retrieve the number of input and output tokens of `CypherAgent'. However, the challenge is identifying the corresponding fields within `CypherAgent' as this information is not provided in the user's input.

\begin{figure}[h]
    \centering
    \includegraphics[width=0.85\textwidth]{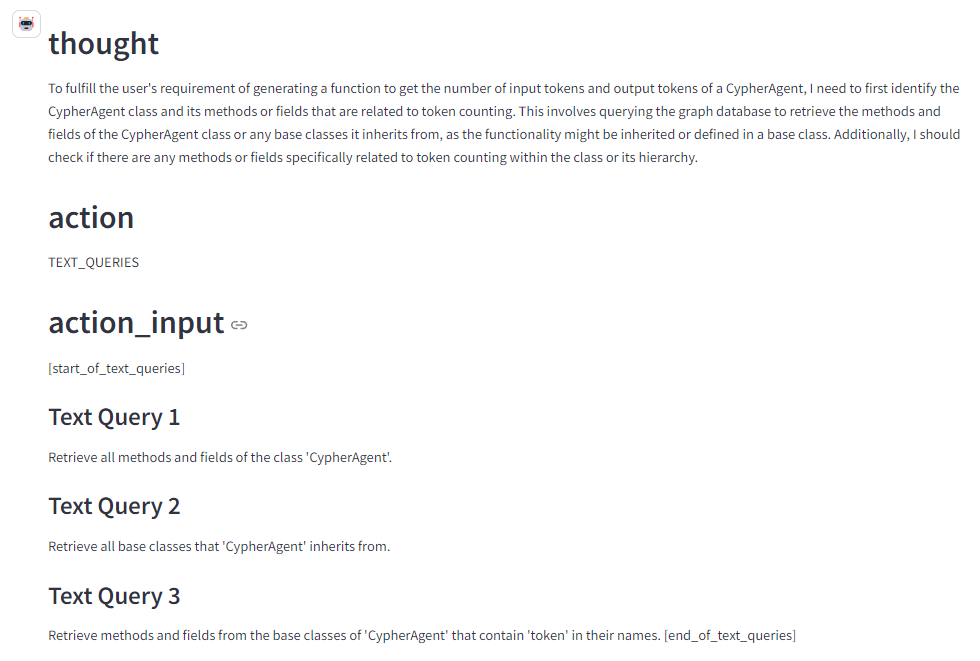}
    \caption{The thought process in determining how to identify the relevant fields.}
\end{figure}

\begin{figure}[h]
    \centering
    \includegraphics[width=0.85\textwidth]{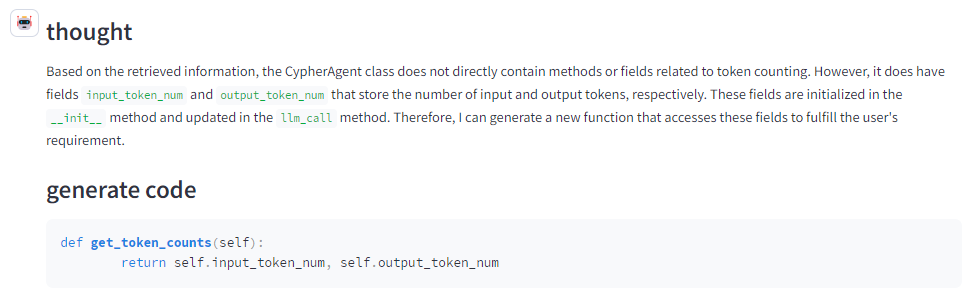}
    \caption{By using Cypher queries, it was discovered that the corresponding fields are `input\_token\_num' and `output\_token\_num', which enables the generation of the correct code.}
\end{figure}

\subsubsection{Example of Code Commentor}

The Code Commentor analyzes code to provide detailed comments, enhancing code readability and maintainability. It leverages the code graph database to understand the code's structure and behavior, ensuring accurate and informative comments.

\begin{figure}[h]
    \centering
    \includegraphics[width=0.8\textwidth]{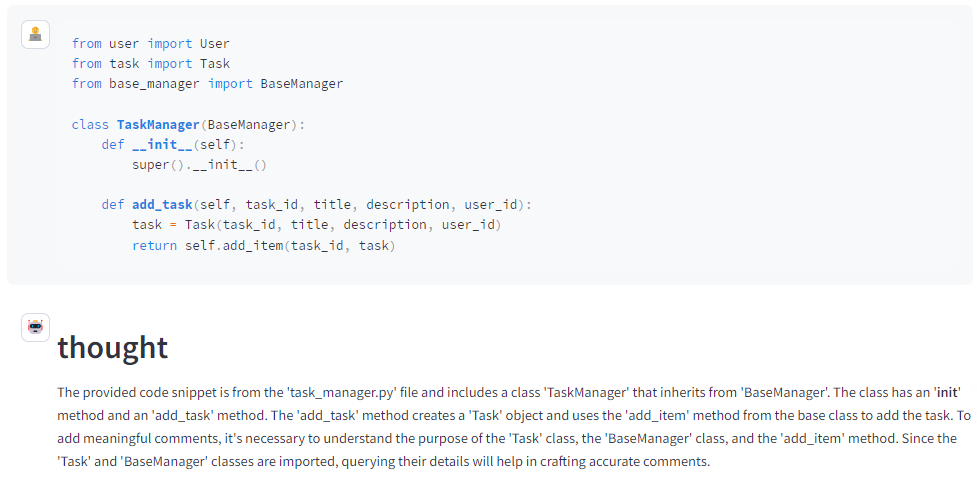}
    \caption{The thought process: Understand the `Task' class and `add\_item' method.}
\end{figure}

\begin{figure}[h]
    \centering
    \includegraphics[width=0.8\textwidth]{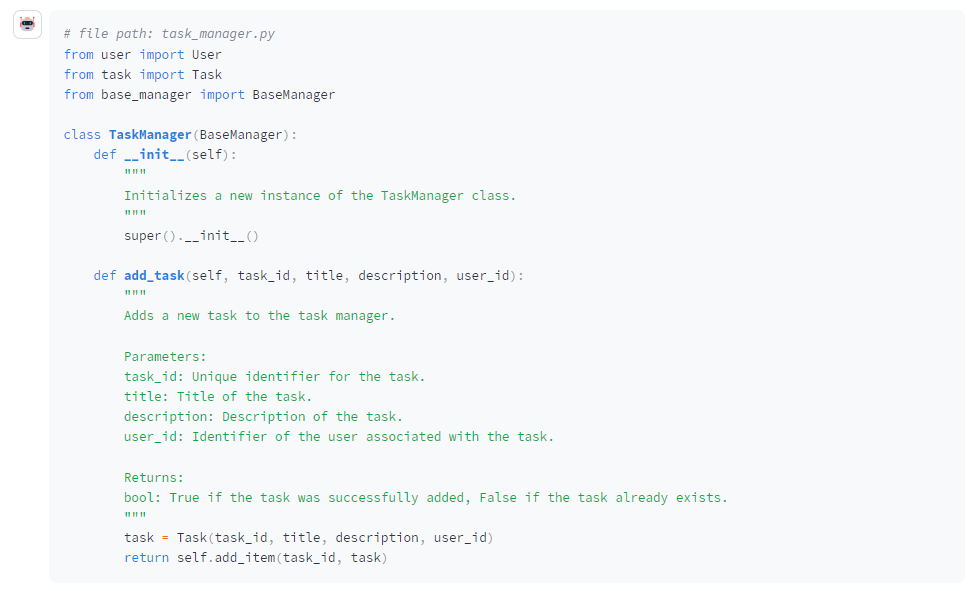}
    \caption{By using Cypher queries, the specific implementation of the return function was obtained, and the return type was clarified.}
\end{figure}

\end{document}

%% file: math_commands.tex
\usepackage{amsmath,amsfonts,bm}

\def\eqref#1{equation~\ref{#1}}
\def\1{\bm{1}}

\DeclareMathAlphabet{\mathsfit}{\encodingdefault}{\sfdefault}{m}{sl}
\SetMathAlphabet{\mathsfit}{bold}{\encodingdefault}{\sfdefault}{bx}{n}